\documentclass[aps,prb,twocolumn,groupedaddress,showpacs]{revtex4-1}

\usepackage{epsfig,amssymb,amsmath}
\usepackage{graphicx}
\usepackage{float}
\usepackage{subfig}

\floatstyle{boxed} 
\restylefloat{figure}

\begin{document}

\title{Joint Density-Functional Theory of the Electrode-Electrolyte Interface: Application to Fixed Electrode Potentials, Interfacial Capacitances, and Potentials of Zero Charge}

\author{Kendra~Letchworth-Weaver}
\author{T.A.~Arias}

\affiliation{Laboratory of Atomic and Solid State Physics, Cornell University, Ithaca, NY 14853}

\date{\today}

\begin{abstract}
This work explores the use of joint density-functional theory, a new form of density-functional theory for the {\em ab initio} description of electronic systems in thermodynamic equilibrium with a liquid environment, to describe electrochemical systems.  After reviewing the physics of the underlying fundamental electrochemical concepts, we identify the mapping between commonly measured electrochemical observables and microscopically computable quantities within an, in principle, exact theoretical framework.  We then introduce a simple, computationally efficient approximate functional which we find to be quite successful in capturing {\em a priori} basic electrochemical
phenomena, including the capacitive Stern and diffusive Gouy-Chapman regions in the electrochemical double layer, quantitative values for interfacial capacitance, and electrochemical potentials of zero charge for a series of metals. We explore surface charging with applied potential and are able to place our {\em ab initio} results directly on the scale associated with the Standard Hydrogen Electrode (SHE).  Finally, we provide explicit details for implementation within standard density-functional theory software packages at negligible computational cost over standard calculations carried out within vacuum environments.
\end{abstract}


\maketitle

\section{INTRODUCTION}

{\em Ab initio} calculations have shed light on many questions in
physics, chemistry, and materials science, including chemical
reactions in solution \cite{SprikBlumberger05,SprikBlumberger06} and at surfaces.\cite{Radeke97,Norskov-Greeley02,Gross02}  However, first principles calculations have offered less insight in the complex and multi-faceted field of electrochemistry, despite the potential scientific and technological impact of advances in this field.  Because the fundamental microscopic mechanisms involved in oxidation and reduction at electrode surfaces are often unknown and are difficult to determine experimentally, \cite{Shi06} rich scientific opportunities are available for theoretical study.  From a technological perspective, practicable first principles calculations could become a vital tool to direct the experimental search for better catalysts with significant potential societal impact: as just one example, economically viable replacement of gasoline powered engines with fuel cells in personal transport systems requires systems operating at a cost of \$35/kW, whereas the current cost is \$294/kW, \cite{FuelCellAnalysis} due mostly to the expense of platinum-based catalyst materials.

The primary challenge which distinguishes theoretical study of electrochemical systems is that including the liquid electrolyte, which critically influences the functioning of the electrochemical cell, requires detailed thermodynamic sampling of all possible internal molecular configurations of the fluid.  Such critical influences include (a) screening of charged systems, (b) establishment of an absolute potential reference for oxidation and reduction potentials, and (c) voltage-dependence of fundamental microscopic processes, including the nature of reaction pathways and transition states.  While there have been attempts at the full {\em ab initio} molecular dynamics approach to this challenge,\cite{Gross09,Gross10,SprikBlumberger05} such calculations are necessarily of the heroic type, require tremendous computational resources, and do not lend themselves to systematic studies of multiple reactions within a series of many candidate systems.  Such studies require development of an alternate approach to first-principles study of electrochemistry.

\subsection{Previous approaches}

One response to the aforementioned challenges is to avoid the issue and lessen the computational cost either by forgoing electronic structure calculation entirely or by neglecting the thermodynamic sampling of the environment.  Some studies have employed classical molecular dynamics with interatomic potentials; \cite{Chandler09,Pounds} however, such semi-empirical techniques often perform poorly when describing chemical reactions involving electron-transfer, which are central to oxidation and reduction reactions.

The latter approach -- single configuration {\em ab initio} calculations -- neglects key phenomena associated with the presence of an electrolyte liquid in equilibrium. The most direct single configuration {\em ab initio} approach pursued to date is to study the relevant reactions on a surface in vacuum and to study trends and correlations with the behavior in electrochemical systems.\cite{Norskov04, Norskov2009NatChem}  Some of these studies are done in a constant charge or constant potential ensemble\cite{Lozovoi03} to allow variation of the applied electrode potential.  This approach, however, does not include critical physical effects of the electrolyte such as the dielectric response of the liquid environment and the presence of high concentrations of ions in the supporting electrolyte.  In response, an intermediate approach is to include a layer or few layers of explicit water molecules into the calculation.\cite{Neurock06,Neurock07,KarlsbergNorskov07,Rossmeisl11}  

Such an approach is problematic for a number of reasons.  First, actual electrochemical systems can have rather long ionic screening lengths (30 \AA~ for an ionic concentration of 0.01 M), which would require large amounts of explicit water.  Second, simulation of the actual effects of dipolar and ionic screening in the fluid requires extensive sampling of phase space, corresponding to very long run times. Indeed, in some references, only one layer of frozen water without thermal or time sampling is included. \cite{Norskov07}  Moreover, as most reactions of interest occur at potentials away from the potential of zero charge, such calculations must include a net charge, which can be problematic in typical solid-state periodic supercell calculations. One may compensate for this charge with a uniform charged background extending throughout the unit cell, both the liquid and the solid regions,\cite{Neurock07} but this distribution does not reflect the electrochemical reality.  Other methods include an explicit reference electrode with a corresponding negative surface charge to keep the unit cell neutral,\cite{Lozovoi03} but this requires a somewhat arbitrary choice of where to place the compensating electrode and may not lead to realistic potential profiles. More recently, modeling the electrolyte by a layer of explicit hydrogen atoms was shown to provide a source of electrons for charged surface calculations while keeping the unit cell neutral.\cite{Norskov10}  Again, however, this approach requires either judicious choice of the locations of the corresponding protons which make up the corresponding reference electrode or computationally intensive thermodynamic sampling.

Another broad approach constructs an approximate {\em a posteriori} continuum model\cite{GygiFattebert} for both the dielectric response of the water molecules and the Debye screening effects of the ions and performs {\em ab initio} calculations where the electrostatic potential is determined by solving Poisson-Boltzmann-like equations.\cite{Otani06,AndersonPRB08,MarzariDabo}  Explicit inclusion of a few layers of explicit water molecules and ionic species within the {\em ab initio} calculations can further enhance the reliability of this approach without dramatic additional computational cost. While including explicitly the most recognized physical effects of the electrolyte, such Poisson-Boltzmann-like approaches do not arise from an exact underlying theory. Thus, they may disregard physically relevant effects, such as the non-locality and non-linearity of the dielectric response of liquid water and the surface tension associated with formation of the liquid-solid interface.  We note, for instance, that a typical electrochemical field strength would be a 0.1 V drop over a double layer width of 3 Angstroms, or 300 MV/m, a field at which the bulk dielectric constant of water is reduced by about one-third, strongly indicating that non-linear dielectric saturation effects are present in actual electrochemical systems, particularly near the liquid-solid interface, and ultimately should be captured naturally for an {\em ab initio} theory to be truly predictive and reliable.   

\subsection{Joint density-functional theory approach}

This work begins by placing the aforementioned modified Poisson-Boltzmann approaches on a firm theoretical footing within an, {\em in principle}, exact density-functional theory formalism, and then describes the path to including all of the aforementioned effects in a fully rigorous {\em ab initio} density functional.  The work then goes on to elucidate the fundamental physics underlying electrochemistry and provide techniques for computation of fundamental electrochemical quantities from a formal perspective.  The work then shifts focus and introduces an extremely simplified functional for initial exploration of the potential of our overall approach for practical calculations.  The equations which result at this high level of simplification resemble those introduced by others\cite{AndersonPRB08,MarzariDabo} from an {\em a posteriori} perspective, thus putting those works on a firmer theoretical footing and showing them in context as approximate versions of a rigorous underlying approach.  We then work within this simplified framework to explore -- in more depth than previously in the literature -- fundamental physical effects in electrochemistry, including the microscopic behavior of the electrostatic potential near an electrode surface, the structure of the electrochemical double layer, differential capacitances, and potentials of zero charge across a series of metals.  The encouraging results which we obtain even with this highly simplified functional indicate that the overall framework is sound for the exploration of physical electrochemical phenomena and strongly suggests that the more accurate functionals under present development \cite{Lischner10} will yield accurate, fully {\em ab initio} results.

Section~II begins by laying out our theoretical framework, Section~III describes connections between experimental electrochemical observables and microscopic {\em ab initio} computables. Section~IV introduces a simple approximate functional which offers a computationally efficient means of bridging connections to experimental electrochemistry. Section V provides specific details about electronic structure calculations of transition metal surfaces. Finally, Section~VI presents electrochemical results for those metallic surfaces obtained with our simplified functional and Section~VII concludes the paper. The appendices include technical information regarding implementation of our functional within a pseudopotential framework.

\section{THEORETICAL FRAMEWORK}

As described in the Introduction, much of the challenge in performing
realistic {\em ab initio} electrochemistry calculations comes not only from the need to include explicitly the atoms composing the environment but also from the need to perform thermodynamic averaging over the locations of those atoms.  Recently, however, it was proved rigorously that one can compute exact free-energies by including the environment in a joint density-functional theory framework.
\cite{Petrosyan05,Petrosyan07}  Specifically, this previous work shows that the free energy $A$ of an explicit quantum mechanical system with its nuclei at fixed locations while in thermodynamic equilibrium with a liquid environment (including full quantum mechanical treatment of the environment electrons and nuclei), can be obtained by the following variational principle, \cite{Petrosyan07}
\begin{flalign}
A&=\min_{n(r),\{N_{\alpha}(r)\}}\{G[n(r),\{N_{\alpha}(r)\},V(r)
]\nonumber\\&-\int d^3r V(r)n(r)\}
\label{FullFunc}
\end{flalign}
where $G[n(r),\{N_\alpha(r)\},V(r)]$ is a universal functional of the
electron density of the explicit system $n(r)$, the densities of the
nuclei of the various atomic species in the environment $\{N_\alpha(r)\}$, and the electrostatic potential from the nuclei of the explicit system $V(r)$.  The functional $G[n(r),\{N_\alpha(r)\},V(r)]$ is universal in the sense that it
depends only on the nature of the environment and that its dependence on the explicit system is only through the electrostatic potential of the nuclei included in $V(r)$ and the electron density of the explicit system $n(r)$.  With this
functional dependence established, one can then separate the functional into large, known portions and a smaller coupling term ultimately to be approximated,\cite{Petrosyan07}
\begin{flalign}
G[n(r),\{N_\alpha(r)\},V(r)]&\equiv A_{KS}[n(r)]\nonumber
+\Omega_{lq}[\{N_\alpha(r)\}]\\& +\Delta A[n(r),\{N_\alpha(r)\},V(r)] \label{eq:fJDFT}
\end{flalign}
where  $A_{KS}[n(r)]$ and  $\Omega_{lq}[\{N_\alpha(r)\}]$ are, respectively, the standard universal Kohn-Sham electron-density functional of the explicit solute system in isolation (including its nuclei and their interaction with its electrons) and the "classical" density-functional for the liquid solvent environment in isolation.  The remainder, $\Delta A[n(r),\{N_\alpha(r)\},V(r)]$ is then the coupling term between the solute and solvent.

For $A_{KS}[n(r)]$, one can employ any of the popular approximations to electronic density functional theory such as the local-density approximation (LDA), or more sophisticated functionals such as the generalized-gradient approximation (GGA). \cite{PW91} On the other hand, functionals $\Omega_{lq}[\{N_\alpha(r)\}]$ for liquid solvents such as water are generally less-well developed, though the field has progressed significantly over the past few years.  For example, one recent, numerically efficient functional for liquid water reproduces many of the important factors determining the interaction between the liquid and a solute, including the linear {\em and nonlinear} non-local dielectric response, the experimental site-site correlation functions, the surface tension, the bulk modulus of the liquid and the variation of this modulus with pressure, the density of the liquid and the vapor phase, and liquid-vapor coexistence \cite{Lischner10}.  A framework employing such a functional would be more reliable than the modified Poisson-Boltzmann approaches available to date, which do not incorporate any of these effects except for the linear local dielectric response appropriate to macroscopic fields.  Inclusion of the densities of any ions in the electrolyte environment among the $\{N_\alpha(r)\}$ is a natural way to include their effects into $\Omega_{lq}[\{N_\alpha(r)\}]$ and provide ionic screening into the overall framework.

Finally, developing approximate forms for the coupling $\Delta A[n(r),\{N_\alpha(r)\},V(r)]$ in (\ref{eq:fJDFT}) remains an open area of research.  In an early attempt, Petrosyan and co-workers\cite{Petrosyan07} employed a simplified $\Omega_{lq}[\{N_\alpha(r)\}]$ using a single density field $N(r)$ to describe the fluid.  In that preliminary work, because such an $N(r)$ gives no explicit sense of the orientation of the liquid molecules, the tendency of these molecules to orient and screen long-range electric fields was included {\em a posteriori} into a simplified linear (but nonlocal) response function.  In a more complete framework with explicit distributions for the oxygen and hydrogen sites among the $\{N_\alpha(r)\}$ the full non-local and non-linear dielectric response can be handled completely {\em a priori}.\cite{Lischner10}  Beyond long-range screening effects, the coupling $\Delta A[n(r),\{N_\alpha(r)\},V(r)]$ must also include effects from direct contact between the solvent molecules and the solute electrons.  Because the overlap between the molecular and electron densities is small, the lowest-order coupling, very similar to the ``molecular'' pseudopotentials of the type introduced by Kim {\em et al.},\cite{Cho96} would be a reasonable starting point. Using such a pseudopotential approach (with only the densities of the oxygen atoms of the water molecules), Petrosyan and coworkers \cite{Petrosyan07} obtained good agreement (2 kcal/mole) with experimental solvation energies, without any fitting of parameters to solvation data.  Combining a coupling functional $\Delta A$ similar to that of Petrosyan and coworkers with more explicit functionals $\Omega_{lq}[\{N_\alpha(r)\}]$ for the liquid\cite{Lischner10} and standard electron density functionals $A_{KS}[n(r)]$ for the electrons, is thus a quite promising pathway to highly accurate {\em ab initio} description of systems in equilibrium with an electrolyte environment.

\section{CONNECTIONS TO ELECTROCHEMISTRY}

Turning now to the topic of electrochemistry, we present a general theoretical framework to relate the results of {\em ab initio} calculations to experimentally measurable quantities, beginning with a brief review of the electrochemical concepts.

\subsection{Electrochemical potential}

In the electrochemical literature, the {\em electrochemical potential} $\bar{\mu}$ of the electrons in a given electrode is defined as the energy required to move electrons from a reference reservoir to the working electrode. This potential is often conceptualized as a sum of two terms, $\bar{\mu}=\mu_{int}-F\Phi$, where $\mu_{int}$ is the purely ”chemical” potential (due to concentration gradient and temperature,chemical bonding, etc.), $\Phi$ is the external, macroscopic electrostatic potential, and F is Faraday's constant. (Note that $F=N_A e$ has the numerical value of unity in atomic units, where chemical potentials are measured {\em per particle} rather than {\em per mole}.)  In the physics literature, this definition for $\bar{\mu}$ (when measured {\em per particle}) corresponds precisely to the ``chemical potential for electrons,'' which appears for instance in the Fermi occupancy function $f=[e^{(\epsilon-\bar{\mu})/k_B T}+1]^{-1}$.

\subsection{Electrode potential}

In a simple, two-electrode electrochemical cell, the driving force for chemical reactions occurring at the electrode surface is a voltage applied between the reference electrode and working electrode.  In the electrochemical literature, this voltage is known as the {\em electrode potential} ${\cal E}$, defined as the electromotive force applied to a cell consisting of a working electrode and a reference electrode. In atomic units (where the charge of an electron is unity), the electrode potential is thus equivalent to the energy (per fundamental charge $e$) supplied to transfer charge (generally in the form of electrons) from the reference to the working electrode, assuming no dissipative losses.  Under conditions where diffusion of molecules and reactions occurring in the solution are minimal, this energy is completely transferred to the electrons in the system, causing a corresponding change in the electrochemical potential of the electrons in the working electrode.

\begin{figure}
\centering

\subfloat[]{\label{fig:1a}\includegraphics[width=5cm]{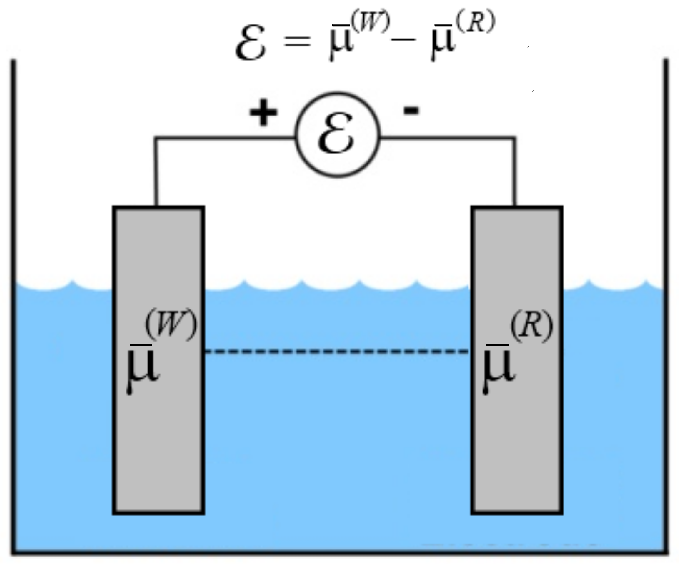}}

\subfloat[]{\label{fig:1b}\includegraphics[width=7.5cm]{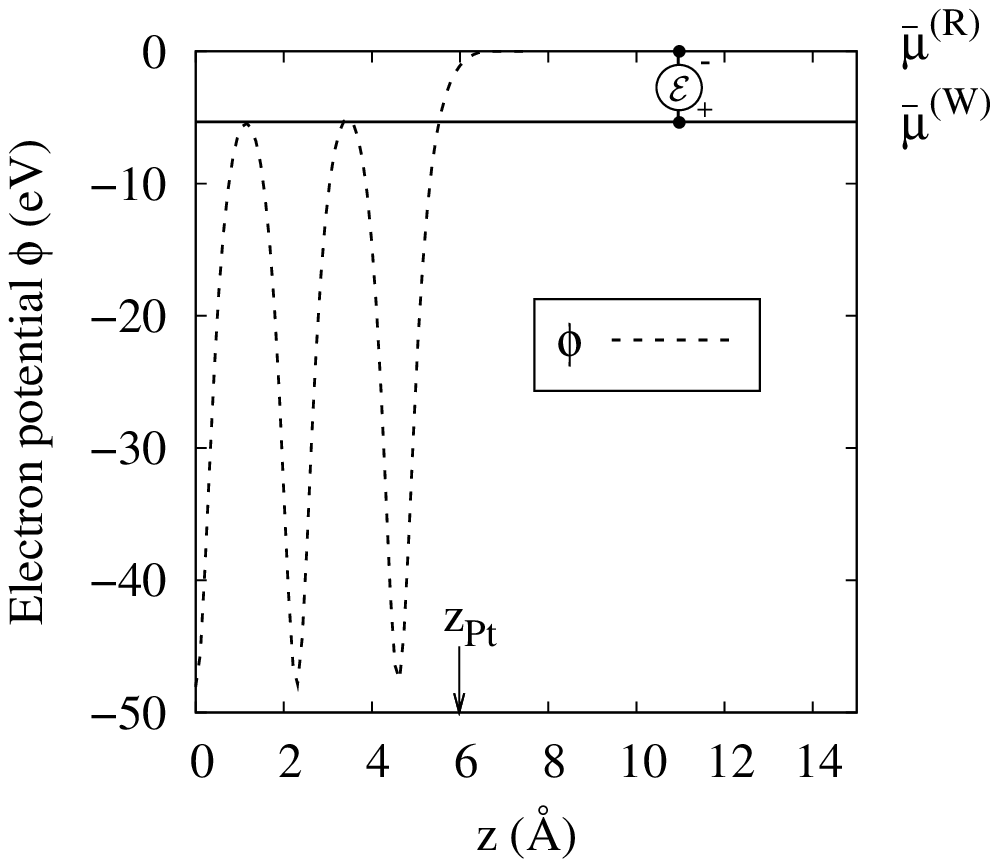}}

\caption{(a) Schematic of an electrochemical cell.  The working electrode is explicitly modeled while the reference electrode is fixed at zero. (b) Relationship between the microscopic electron potential $\langle\phi(z)\rangle$ (averaged over the directions parallel to the surface), electrochemical potential, and applied potential for a Pt (111) surface. The large variations in potential to the left of $z_{Pt}$ correspond to the electrons and ionic cores comprising the metal while the decay into the fluid region is visible to the right of $z_{Pt}$.}
\label{Figure1}
\end{figure}

An idealized two-terminal electrochemical cell controls the chemical potential of a working electrode $\bar{\mu}^{(W)}$ through application of an electrode potential ${\cal E}$ (voltage) between it and a reference electrode of known chemical potential $\bar{\mu}^{(R)}$ (See Figure~\ref{Figure1}(a)).  With the application of the electrode potential ${\cal E}$, the energy {\em cost} to the electrochemical cell, under reversible (lossless) conditions, to move a single electron from the reference electrode to the working electrode is $dU=-\bar{\mu}^{(R)}+ \bar{\mu}^{(W)}+{\cal E}$.  Here, the electrode potential appears with a positive sign, because to move a negative charge from the negative to positive terminal requires a net investment of energy, and thus {\em cost} to the electrochemical cell, against the source of the potential ${\cal E}$.  Under equilibrium conditions, we must have $dU=0$, so that ${\cal E}=\bar{\mu}^{(R)}-\bar{\mu}^{(W)}$.

As Section~IV shows, the electrostatic model which we employ for ionic screening in this work establishes a fixed reference such that the microscopic {\em electron} potential $\phi$ (the Coulomb potential energy of an electron at a given point) is zero deep in the liquid environment far from the electrode (See Figure~\ref{Figure1}(b)) -- implying that the macroscopic {\em electrostatic} potential $\Phi$ (which differs in overall sign from $\phi$) there is also zero. A convenient working electrode thus corresponds to electrons solvated deep in the fluid, which will have electrochemical potential $\bar{\mu}^{(R)}=\mu_{int}^{(s)}-F \Phi = \mu_{int}^{(s)}$, where $\mu_{int}^{(s)}$ corresponds to the solvation energy of an electron in the liquid.  Referring the scale of the electrochemical potential to such solvated electrons (so that $\mu_{int}^{(s)}\equiv 0$), we then have $\bar{\mu}^{(R)}=0$, so that ${\cal E}=-\bar{\mu}^{(W)}$.  In sum, the opposite of the electronic chemical potential in our {\em ab initio} calculations corresponds precisely to the electrode potential relative to solvated electrons. In practice, the choice of approximate density functionals $\Omega_{lq}[\{N_\alpha(r)\}]$ and  $\Delta A[n(r),\{N_\alpha(r)\},V(r)]$ sets the value of the electron solvation energy; each model fluid corresponds to a different reference electrode of solvated electrons.  Section~VII demonstrates the establishment of the electrochemical potential of such a model reference electrode relative to the standard hydrogen electrode (SHE).

\subsection{Potential of zero charge (PZC) and differential capacitance.}

For any given working electrode, a specific number of electrons, and thus electronic chemical potential $\bar{\mu}$, is required to keep the system electrically neutral.  The corresponding electrode potential (${\cal E}=-\bar{\mu}$) is known as the potential of zero charge.  Adsorbed ions from the electrolyte or other contaminants on the electrode surface create uncertainty in the experimental determination of the potential of zero charge.  One advantage to {\em ab initio} calculation is the ability to separate the contribution due to adsorbed species from the contribution of the electrochemical double layer, the latter being defined as the potential of zero free charge (PZFC).  Experimentally, only the potential of zero total charge (PZTC), which includes the effects of surface coverage, may be measured directly, and the potential of zero free charge can only be inferred.\cite{Cuesta}  {\em Ab initio} approaches such as ours allow for the possibility of controlled addition of adsorbed species and direct study of these issues.

At other values of the electrode potential ${\cal E}$, the system develops a charge per unit surface area $\sigma \equiv Q/A$.  From the relationship between these two quantities $\sigma({\cal E})$, one can then determine the differential capacitance per unit area ${\cal C} \equiv \frac{d\sigma}{d{\cal E}}$.  The total differential capacitance of a metal is determined by both the density of states of the metal surface ${\cal C}_{\mbox{DOS}}$, also known as the quantum capacitance,\cite{QCap} and the capacitance associated with the fluid ${\cal C}_{\mbox{fl}}$. These capacitances act in series, so that full differential capacitance is given by
\begin{flalign}
{\cal C}^{-1}={\cal C}_{\mbox{fl}}^{-1}+{\cal C}_{\mbox{DOS}}^{-1}.
\label{Capseries}
\end{flalign}
In typical systems, $C_{\mbox{DOS}} \sim 100-1000 \mu F/cm^2$ is larger than the fluid capacitance (typically ${\cal C}_{\mbox{fl}} \sim 15-100 \mu F/cm^2$), so when the two are placed in series, the fluid capacitance dominates.

The fluid capacitance may be further decomposed into two capacitors acting in series,
\begin{flalign}
{\cal C}_{\mbox{fl}}^{-1}={\cal C}_{\Delta}^{-1}+{\cal C}_{\kappa}^{-1},
\label{CapFl}
\end{flalign}
as in the Gouy-Chapman-Stern model for the electrochemical double layer.\cite{Gouy,Chapman,Stern} The surface charge on the electrode and the first layer of oppositely charged ions behave like a parallel plate capacitor with distance $\Delta$ between the plates. $\Delta$ indicates the distance from the electrode surface to the first layer of ions -- called the outer Helmholtz layer for non-adsorbing electrolytes.  The capacitance per unit area for this simple model is ${\cal C}_{\Delta}=\frac{\epsilon_0}{\Delta}$, analogous to the Helmholtz capacitance. For a gap size $\Delta \sim 0.5~$\AA, this model leads to a ``gap'' capacitance of about 20 $\mu F/cm^2$.  Additional capacitance arises from the diffuse ions in the liquid. where the model for this capacitance ${\cal C}_{\kappa}=\epsilon_b\epsilon_0\kappa \mbox{cosh}(\frac{e\phi(\Delta)}{2k_BT})$ is also well-known from the electrochemistry literature.\cite{BardFaulkner}  In the limit where most of the voltage drop is found in the outer Helmholtz layer ($\phi(\Delta)\sim k_BT$), this expression reduces to a constant value which depends only on the concentration of ions in the electrolyte and the bulk dielectric constant of the fluid $\epsilon_b$: ${\cal C}_{\kappa}=\frac{\epsilon_b\epsilon_0}{\kappa^{-1}}$. For water with a 1.0 M ionic concentration, the ``ion'' capacitance is ${\cal C}_{\kappa}=$240 $\mu F/cm^2$, an order of magnitude larger than the ``gap'' capacitance. At this high ionic concentration, the ``gap'' (Helmholtz) capacitance dominates not only the fluid capacitance, but also the total capacitance.  For lower concentrations of ions, the magnitude of the ``ion'' capacitance becomes more comparable to the ``gap'' capacitance and voltage-dependent nonlinear effects in the fluid could become important.

\subsection{Cyclic voltammetry}

A powerful technique for electrochemical analysis is the cyclic voltammogram, in which current is measured as a function of voltage swept cyclically at a constant rate.  Such data yield detailed information about electron transfer in complicated electrode reactions, with sharp peaks corresponding to oxidation or reduction potentials for chemical reactions taking place at the electrode surface.  Because current is a time-varying quantity and density-functional theory does not include information about time dependence and reaction rates, careful reasoning must be employed to compare {\em ab initio} calculations to experimental current-potential curves. Previous work has correlated surface coverage of adsorbed hydrogen with current in order to predict cyclic voltammograms for hydrogen evolution on platinum electrodes.\cite{KarlsbergNorskov07}  This simple model for a cyclic voltammogram is intrinsically limited at a full monolayer of hydrogen adsorption, rather than by the more realistic presence of mass transport and diffusion effects, but nonetheless provides useful comparisons to experimental data.

Using a similar approach, our framework gives the predicted current density $J$ directly through the chain rule as
$$J=\frac{d\sigma}{dt}=\frac{d{\cal E}}{dt} \frac{d\sigma}{d{\cal E}} \equiv K {\cal C}({\cal E}),$$ where $K=\frac{d\cal E}{dt}$ is the voltage sweep rate, and ${\cal C}({\cal E})$ is the differential capacitance per unit area at electrode potential ${\cal E}$, as defined above. For the bare metal surfaces with no adsorbates studied in Section VII of this work, only the double layer region structure is visible, but the technique may be generalized to study chemical reactions at the electrode surface. The current density curve is simply proportional to the differential capacitance per unit area ${\cal C}$ as long as the state of the system varies adiabatically and the voltage sweep rate is significantly slower than the reaction rate. In the adiabatic limit, features in the charge-potential curves calculated for reaction intermediates and transition states can be compared directly with peaks in cyclic voltammograms to predict oxidation and reduction potentials from first principles. 

\section{IMPLICIT SOLVENT MODELS}

For computational expediency and to explore the performance of the overall framework for quantities of electrochemical interest, we now introduce a highly approximate functional.  Despite its simplicity, we find that the model below leads to very promising results for a number of physical quantities of direct interest in electrochemical systems. The first step in this approximation is to minimize with respect to the liquid nuclear density fields in the fully rigorous functional \cite{Petrosyan05} so that Eq. (\ref{FullFunc}) becomes
\begin{flalign}
\tilde{A}&=\min_{n(r)}(A_{KS}[n(r),\{Z_I,R_I\}]\nonumber\\&+\Delta \tilde{
A}[n(r),\{Z_I,R_I\}]),
\end{flalign}
with the effects of the liquid environment all appearing in the new term
\begin{flalign}
\Delta \tilde{A}[n(r),\{Z_I,R_I\}] &\equiv
\min_{N_{\alpha}(r)}(\Omega_{lq}[N_\alpha(r)]\nonumber\\& +\Delta
A[n(r),N_\alpha(r),\{Z_I,R_I\}]),
\end{flalign}
where $Z_I$ and $R_I$ are the charges and positions of the surface nuclei (and those of any explicitly included adsorbed species).  This minimization process leaves a functional in terms of {\em only} the properties of the explicit system and incorporates all of the solvent effects implicitly.  Up to this point, this theory is in principle exact, although the exact form of $\Delta \tilde{A}[n(r),\{Z_I,R_I\}]$ is unknown.  For practical calculations this functional must be approximated in a way which captures the underlying physics with sufficient accuracy.

 \subsection{Approximate functional}

In this initial work, we assume that the important interactions between the solvent environment and the explicit solute electronic system are all electrostatic in nature.  Our rationale for this choice is the fact that most electrochemical processes are driven by (a) the surface charge on the electrode and the screening due to the dielectric response of the liquid solvent and (b) the rearrangement of ions in the supporting electrolyte.  To incorporate these effects, we calculate the {\em electron} potential  $\phi(r)$ (the Coulomb potential energy of an electron at a given point, which equals $-e$ times the {\em electrostatic} potential) due to the electronic and atomic core charges of the electrode and couple this potential to a spatially local and linear description of the liquid electrolyte environment, yielding
\begin{flalign}
\tilde{A}[n(r),\phi(r)]&= A_{TXC}[n(r)]\nonumber\\
& +\int d^3r\{\phi(r)\left(n(r)-N(r,\{Z_I,R_I\})\right)\nonumber\\
& -\frac{\epsilon(r)}{8\pi}|\nabla\phi(r)|^2-\frac{\epsilon_b\kappa^2(r)}{8\pi}(\phi(r))^2\},
\label{ApproxFunc}
\end{flalign}
where $A_{TXC}[n(r)]$ is the Kohn-Sham single-particle kinetic plus
exchange correlation energy, $n(r)$ is the full electron density of the explicit system (including both core and valence electrons), and $N(r,\{R_I,Z_I\})$ is the nuclear particle density of the explicit solute system with nuclei of atomic number $Z_I$ at positions $R_I$, $\epsilon_b$ is the bulk dielectric constant of the solvent, and $\epsilon(r)$ and $\kappa(r)$ are local values, respectively, of the dielectric constant and the inverse Debye screening length due to the presence of ions in the fluid.  We emphasize that, despite the compact notation in (\ref{ApproxFunc}), in practice we employ standard Kohn-Sham orbitals to capture the kinetic energy and, as the appendices detail, we employ atomic pseudopotentials rather than direct nuclear potentials, so that $N(r,\{R_I,Z_I\})$ does not consist in practice of a set of Dirac $\delta$-functions.

To determine local values of the quantities $\epsilon(r)$ and $\kappa(r)$ above, we relate them directly to the local average density of the solvent $N_{lq}(r)$ as
\begin{flalign}
\epsilon(r)&\equiv 1+\frac{N_{lq}(r)}{N_b}(\epsilon_b-1)\nonumber\\
\kappa^2(r) &\equiv \kappa_b^2 \frac{N_{lq}(r)}{N_b},
\label{eps-kapp}
\end{flalign}
where $N_b$ and $\epsilon_b$ are, respectively, the bulk liquid number density (molecules per unit volume) and the bulk dielectric constant, and $\kappa_b^2=\frac{e^2}{\epsilon_b\,k_BT}\sum_i N_i Z_i^2$ is the square of the inverse Debye screening length in the bulk fluid, where $Z_i$ and  $N_i=c_i N_A$ are the valences and number densities of the various ionic species.  Finally, our model for the local liquid density depends on the full solute electron density $n(r)$ at each point through the relation
\begin{equation}
N_{lq}(n)\equiv\frac{N_b}{2}\mbox{erfc}\left(\frac{\ln{(n/n_0)}}{\sqrt{2}\gamma}\right),
\label{Nl}
\end{equation}
a form which varies smoothly (with transition width $\gamma$) from the bulk liquid density $N_b$ in the bulk solvent where the electron density from the explicit system is less than a transition value $n_0$ to zero inside the cavity region associated with the solute, defined as those points where $n(r) > n_0$.  This form for $N_{lq}(n)$ reproduces solvation energies of small molecules in water without ionic screening to within 2 kcal/mol,\cite{Petrosyan05} when the parameters in Eq.~(\ref{Nl}) have values $\gamma=0.6$ and $n_0=4.73\times 10^{-3}\ $\AA$^{-3}$.

The stationary point of the functional in Eq.~(\ref{ApproxFunc}) determines the physical state of the system and is actually a saddle point which is a minimum with respect to changes in $n(r)$ (or, equivalently, the Kohn-Sham orbitals) and a maximum with respect to changes in $\phi(r)$.  Setting to zero the variation of Eq.~(\ref{ApproxFunc}) with respect to the single-particle orbitals generates the usual Kohn-Sham, Schr\"{o}dinger-like, single-particle equations with $\phi(r)$ replacing the Hartree and nuclear potentials and results in the modified Poisson-Boltzmann equation,
\begin{flalign}
\nabla \cdot \left( \epsilon(r) \nabla \phi(r) \right) - \epsilon_b\kappa^2(r)\phi(r)
\nonumber\\=-4\pi\left(n(r)-N(r,\{R_I,Z_I\})\right).
\label{mPB}
\end{flalign}
Self-consistent solution of this modified Poisson-Boltzmann equation for $\phi(r)$ along with solution of the corresponding traditional Kohn-Sham equations defines the final equilibrium state of the system.

\begin{figure}
\centering
\subfloat[]{\label{fig:2a}\includegraphics[width=8.6cm]{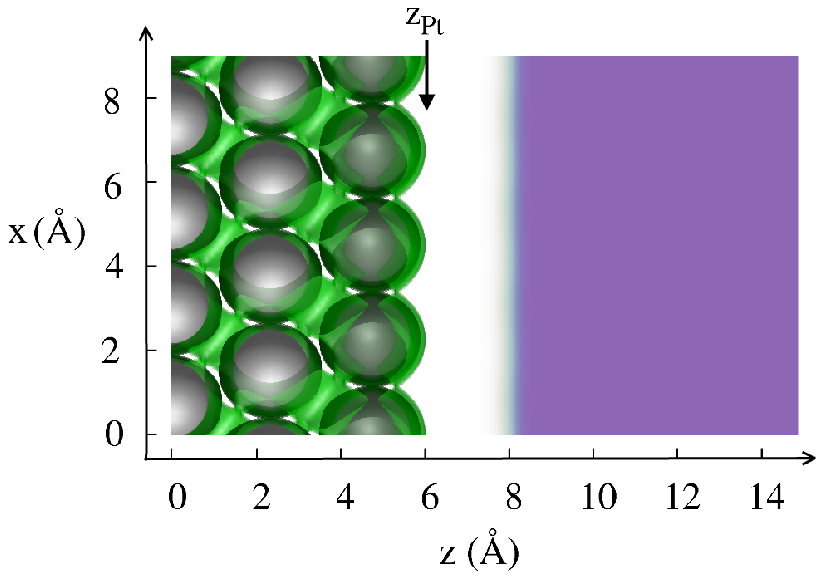}}

\subfloat[]{\label{fig:2b}\includegraphics[width=8.6cm]{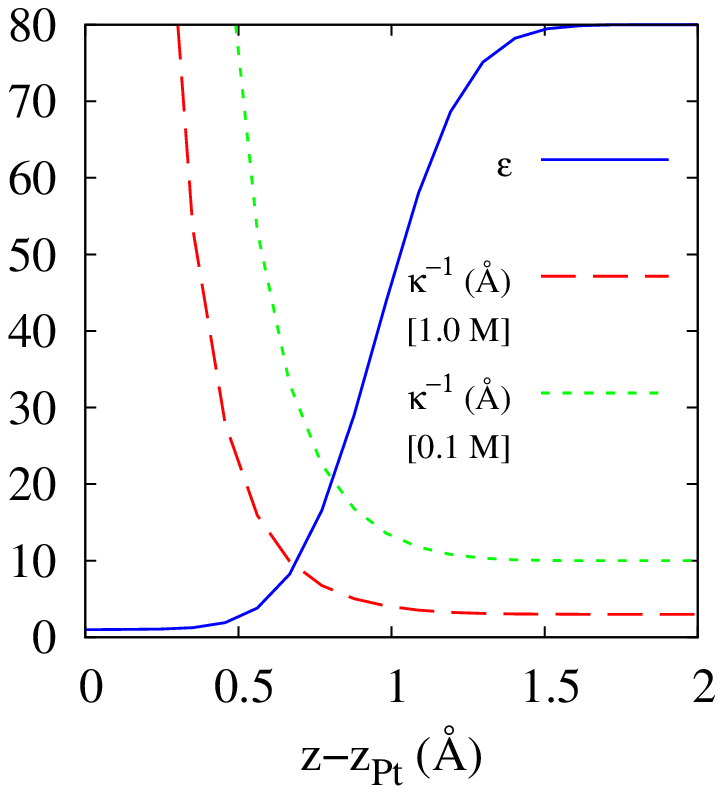}}

\caption{Microscopic and model quantities for Pt(111) surface in
equilibrium with electrolyte: (a) Pt atoms (white), electron density $n(r)$ (green), and fluid density $N_{lq}(r)$ (blue) in a slice passing from surface (left) into the fluid (right) with $z_{Pt}=~5.95~$\AA$~$ indicating the end of the metal, (b) dielectric constant $\langle\epsilon(z)\rangle$ and screening length $\langle\kappa^{-1}(z)\rangle$ (averaged over the planes parallel to the surface) for ionic concentrations of 1.0 M and 0.1 M along a line passing from surface into the fluid. Position $z-z_{Pt}$ measures distance from the end of the metal slab. (See Sections~V~and~VI.)}
\label{Figure2}
\end{figure}

Figure~\ref{Figure2} illustrates the various concepts in this model using actual results from a calculation of the Pt(111) surface, described in Sections~V~and~VI. Figure~\ref{Figure2}(a) shows the electron $n(r)$ and liquid $N_{lq}(r)$ densities in a slice through the system which passes through the metal (left,  $z < z_{Pt}$ ) and the fluid (right,  $z > z_{Pt}$). We define the end of the metal surface $z_{\mbox{metal}}$ by the covalent radius of the last row of metal atoms ($z_{Pt}=5.95~$\AA). The ionic cores and the itinerant valence electrons in the metal are visible, as well as the gap between the surface and the bulk of the fluid. As shown in Figure~\ref{Figure2}(b), the local functions for the dielectric constant $\epsilon(r)$ and the {\em inverse} Debye screening length $\kappa(r)$ respect the correct physical limiting values: $\epsilon_b$ and $\kappa_b$ in the bulk solvent and $\epsilon=1$ and $\kappa=0$ within the surface. The rapid increase in dielectric constant for $0~$\AA$<z-z_{Pt}<1~$\AA$~$ corresponds to the appearance of fluid on the right side and results in the localization of significant charge from the fluid at this location.  The inverse screening length $\kappa$ depends on the concentration of ions in the electrolyte through the bulk liquid value $\kappa_b$.  Figure~\ref{Figure2}(b) shows screening length as a function of distance from the metal surface for both 0.1 and 1.0 molar bulk ionic concentrations. The large screening lengths at positions less than $z_{Pt}$ ensure proper vacuum-like behavior within the metal surface, where all electrons are explicit and thus no implicit screening should appear.

\subsection{Asymptotic behavior of electrostatic potential}

Unlike the standard Poisson equation, which has no unique solution for periodic systems because the zero of potential is an arbitrary constant, the modified Poisson-Boltzmann equation (\ref{mPB}) has a unique solution in periodic systems.  To establish this, we integrate the differential equation (\ref{mPB}) over the unit cell.  The first term, which is the integral of an exact derivative, vanishes.  The remaining terms then give the condition,
\begin{equation}
\int\kappa^2(r)\phi(r)dV=\frac{4\pi}{\epsilon_b}\left(Q_n-Q_N\right),
\end{equation}
where $Q_n$ and $Q_N$ are the total number of electronic and nuclear
charges in the cell, respectively.  Any two $\phi(r)$ which differ by a constant $C$ both can be valid solutions only if $C \int \kappa^2(r) \ dV =0$, so that we must have $C=0$ as long as $\kappa(r)$ is non-zero at any location in the unit
cell.  Thus, any amount of screening at any location in space in the calculation eliminates the usual indeterminacy of $\phi(r)$ by an additive constant, thereby establishing an absolute reference for the zero of the potential.

To establish the nature of this reference potential, we first note
that deep in the fluid, far from the electronic system, the electron
density approaches $n(r)=0$ and the dielectric constant and screening
lengths attain their constant bulk values $\epsilon(r) \rightarrow
\epsilon_b$ and $\kappa(r) \rightarrow \kappa_b$.  Under these
conditions, the Green's function impulse response of (\ref{mPB}) to a unit point charge is
\begin{equation}
\phi(r) = \frac{\exp\left(-\kappa_b\,r\right)}{\epsilon_b\ r},
\label{greensfunction}
\end{equation}
a Coulomb potential screened by the dielectric response of the solvent and exponentially screened by the presence of ions.  Next, we rearrange (\ref{mPB}) so that the left-hand side has the same impulse response as the bulk of the fluid but with a modified source term,
\begin{flalign}
\epsilon_b \nabla^2 \phi(r) - \epsilon_b \kappa_b^2 \phi(r)=\nonumber\\-4\pi\left(\rho_{\mbox{sol}}(r)+\rho_{\mbox{ext}}(r) \right)
\label{mPB-rearrange}
\end{flalign}
where we have defined
\begin{flalign}
\rho_{\mbox{sol}}(r)&\equiv n(r)-N(r)\nonumber\\
\rho_{\mbox{ext}}(r)&\equiv-\frac{1}{4\pi}(\left(\epsilon_b-\epsilon(r)\right)\nabla^2\phi(r)\nonumber\\& -(\nabla
\epsilon(r))\cdot (\nabla \phi(r)) + \epsilon_b
\left(\kappa^2(r)-\kappa_b^2\right) \phi(r)).
\end{flalign}
The key step now is to note that all source terms clearly vanish in the bulk of the fluid where $\rho_{\mbox{sol}}(r) \rightarrow 0$, $\epsilon(r) \rightarrow                 
\epsilon_b = \mbox{constant}$, and $\kappa(r) \rightarrow \kappa_b$. From the exponential decay of the Green's function (\ref{greensfunction}) and the vanishing of
$\rho_{\mbox{sol}}+\rho_{\mbox{ext}}$ in the bulk region of the fluid, we immediately conclude that  $\phi(r) \rightarrow 0$ deep in the
fluid region, thereby establishing that the absolute reference of zero potential corresponds to the energy of an electron solvated deep in the fluid region.

\subsection{Future Improvements}

While offering a computationally efficient and simple way to study electrochemistry, the approximate functional (\ref{ApproxFunc}) is highly simplified and possesses several limitations which the more rigorous approach of Section~II overcomes by coupling of an explicit solvent model for $\Omega_{lq}[N_\alpha(r)]$\cite{Lischner10} to the electronic system through an approach similar to the molecular pseudopotentials proposed by Kim {\em et al.}\cite{Cho96}  Such limitations include the fact that because we employ a linearized Poisson-Boltzmann equation, we do not include the nonlinear dielectric response of the fluid (which other approaches in the literature to date also ignore\cite{AndersonPRB08,MarzariDabo}) or nonlinear saturation effects in the ionic concentrations, both of which become important for potentials greater than a few hundred mV. Despite these limitations, we remain encouraged by the promising results we obtain below for this simple functional and optimistic about the improvements that working within a more rigorous framework would provide.

\section{ELECTRONIC STRUCTURE METHODOLOGY}

All calculations undertaken in this work and presented in Section~VI were all performed within the DFT++ framework \cite{Ismail-Beigi} as implemented in the open-source code JDFTx. \cite{JDFTx} They employed the local-density or generalized-gradient \cite{PW91} approximations using a plane-wave basis within periodic boundary conditions.  The specific materials under study in this paper were platinum, silver, copper, and gold. The (111), (110), and (100) surfaces of each of these metals were computed within a supercell representation with a distance of 10 times the lattice constant of each metal (in all cases around 30 \AA) between surface slabs of thickness of 5 atomic layers.  For these initial calculations, we were very conservative in employing such large regions between slabs to absolutely eliminate electrostatic supercell image effects between slabs.  We strongly suspect that smaller supercells can be used in the future.  All calculations presented employ optimized\cite{Opium} norm-conserving Kleinman-Bylander pseudopotentials\cite{KB} with single non-local projectors in the {\it s}, {\it p}, and {\it d} channels, a plane-wave cutoff energy of 30~H, and employ a $8\times 8\times 1$ $k$-point Monkhorst-Pack\cite{MP} grids to sample the Brillouin zone.

The JDFTx-calculated lattice constants of the bulk metals within both exchange-correlation approximations when using $8 \times 8 \times 8$ $k$-point grids are shown in Table \ref{lattconsts}.
Clearly, the LDA and GGA lattice constants both agree well with the experiment. Except where comparisons are specifically made with LDA results, all calculations in this work employ GGA for exchange and correlation.

\begin{table}[ht]
\caption{ Cubic lattice constant (\AA) in conventional face-centered cubic unit cell}
\centering
\begin{tabular}{c c c c}
\hline\hline
Metal & LDA & GGA & Experiment\cite{CRCHandbook} \\ [0.5ex]
\hline

Pt & 3.93 & 3.94 & 3.92 \\
Cu & 3.55 & 3.67 & 3.61 \\
Ag & 4.07 & 4.13 & 4.09 \\
Au & 4.05 & 4.14 & 4.08 \\ [1ex]

\hline

\end{tabular}
\label{lattconsts}
\end{table}

\section{RESULTS}

To evaluate the promise of our approach, we begin by studying the fundamental behaviors of transition metal surfaces in equilibrium with an electrolyte environment as a function of applied potential.  We find that even our initial highly simplified form of joint density-functional theory reproduces with surprising accuracy a wide range of fundamental physical phenomena related to electrochemistry.  Such transition metal systems, especially platinum, are of electrochemical interest as potential catalysts for both the oxygen reduction reaction (ORR) and th hydrogen evolution reaction (HER).  Molecular dynamics studies of the platinum system in solution, both at the classical \cite{Chandler09} and {\em ab initio} \cite{Gross09,Norskov08} levels, to date have not fully accounted for ionic screening in the electrolyte, which is essential to capturing the complex structure of the electrochemical double layer and the establishment of a consistent reference potential.

For the initial exploratory studies presented in this manuscript, we focus on pristine surfaces without adsorbates in order to establish clearly the relationship between theoretical and experimental quantities and to lay groundwork for future systematic comparison of potential catalyst materials.  Unless otherwise specified, we carry out our calculations with screening lengths of  3~\AA, corresponding to monovalent ionic concentrations of 1.0~M.  We employ these high concentrations because most electrochemical cells include a supporting electrolyte with high ionic concentration chosen to provide strong screening while avoiding (to the extent possible) interaction with and adsorption on the electrode.  Note that, because our present model includes only ionic concentrations and no other species-specific details about the ions in the electrolyte, our results correspond to neutral pH.  Future work will readily explore pH and adsorption effects by including protons and other explicit ions in the electronic-structure portions of the calculation.  One great advantage of the present theoretical approach is the ability to separate the role of the non-adsorbing ions in the supporting electrolyte from the role of the adsorbing ions that interact directly with the surface.

\subsection{Treatment of charged surfaces in periodic boundary conditions}

The application of voltage essential to the {\em ab initio} study of electrochemistry requires precise treatment of charged surfaces not accessible to common electronic structure approaches due to singularities associated with the Coulomb interaction.  In the case of a vacuum environment, the electrostatic potential $\phi(r)$ of even a neutral electrode approaches a physically indeterminate constant which varies with the choice of supercell.  As is well-known, this difficulty compounds radically when a net charge is placed on the surface, resulting in a formally infinite average electrostatic potential in a periodically repeated system.  By default, most electronic structure packages designed for use with periodic systems treat this singularity by setting the $G=0$ Fourier component of $\phi(r)$ to zero, equivalent to incorporating a uniform, neutralizing charge background throughout the region of the computation.  This solution to the Coulomb infinity is not realistic in electrochemical applications where the actual compensating charge appears in the fluid and should not be present in the interior of the electrode.

Another option which has been employed in the electrochemical
context\cite{Otani06} is to include an oppositely charged
counter-electrode located away from the working electrode in the
”vacuum” region of the calculation.  However, including an explicit
density-functional electrode is often computationally prohibitive as
it requires doubling the number of electrons and atoms and requires a
large supercell to prevent image interaction. Implicit inclusion of a
counter-electrode through either coulomb truncation or an external
charge distribution \cite{Otani06} requires an arbitrary choice of the
distribution of external charges representing the counter-electrode,
and such arbitrary choices may result in unphysical electrostatic
potentials, even in the presense of a few explicit layers of neutral
liquid molecules. One realistic choice is to employ Debye
screening as in Eq. \ref{mPB}.  This approach ensures that the long-range decay of $\phi(r)$ into the fluid corresponds to the
behavior of the actual physical system, that the fluid
response contains precisely the correct amount of
compensating charge, and that the potential approaches an
absolute reference, even in a periodic system.

Another more explicit, and hence computationally expensive, option
employed in the electrochemical literature is to add a few layers of
explicit water molecules to the surface and then include explicit
counter-ions (protons) located in the first water
layer.\cite{Rossmeisl11} This approach models some of the most
important effects of
the actual physical distribution of counter-ions, which really should contain
both localized and diffuse components, by considering only the first
layer of localized ions.
 
\begin{figure}

\centering
\subfloat[]{\label{fig:3a}\includegraphics[width=7.5cm]{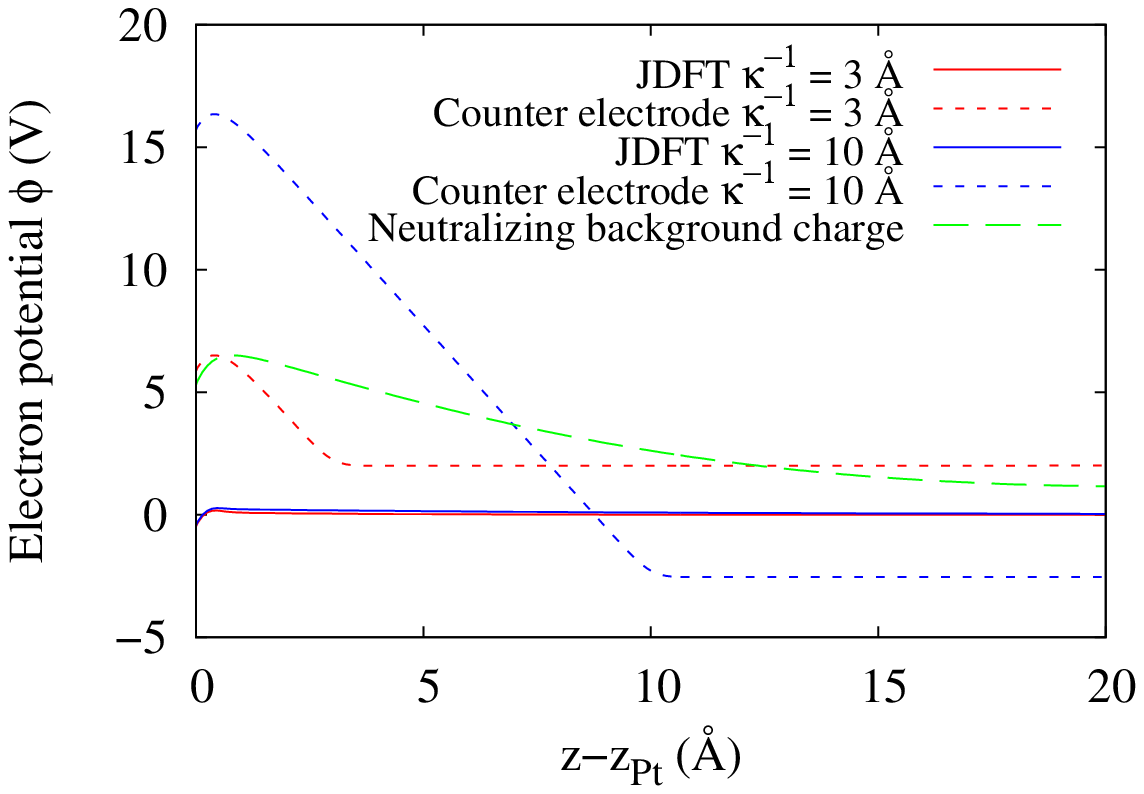}}

\subfloat[]{\label{fig:3b}\includegraphics[width=7.5cm]{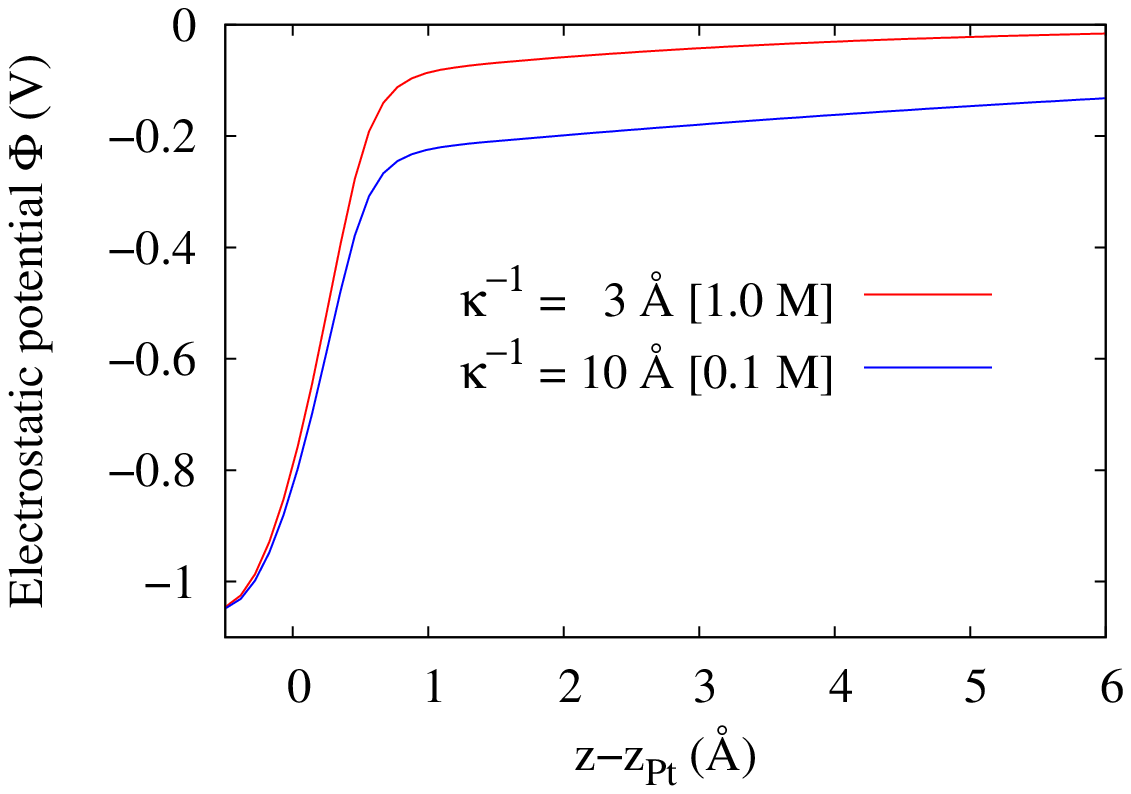}}

\subfloat[]{\label{fig:3c}\includegraphics[width=7cm]{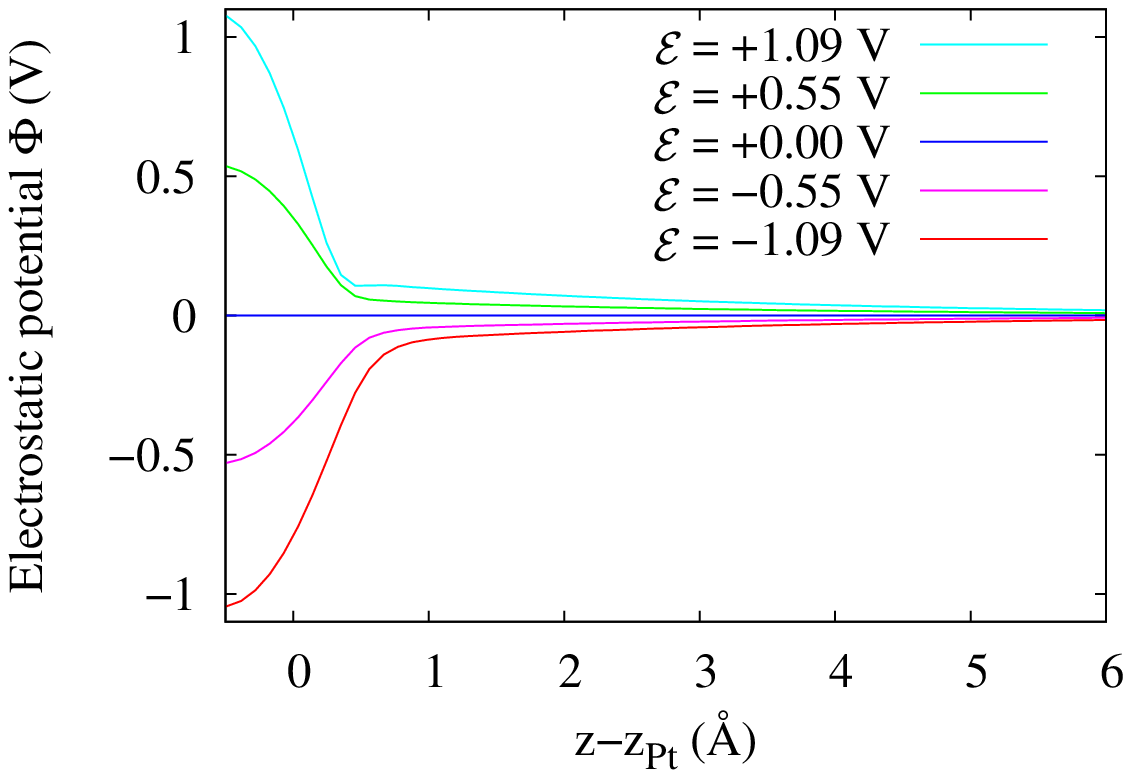}}
 
\caption{Microscopic electron potential energies $\langle \phi(z)\rangle$ and macroscopic electrostatic potentials $\langle\Phi(z)\rangle$ averaged in planes for the Pt (111) surface as a function of distance $z-z_{Pt}$ from the end of the metal surface: (a) $\langle \phi(z)\rangle$ for surface with applied voltage ${\cal E}=-1.09\ \mathrm{V}$ vs. PZC in vacuum (green dashed) and in monovalent electrolytes of $c=1.0\ \mathrm{M}$ (red) and $c=0.1\ \mathrm{M}$ (blue) where the dotted lines represent calculations with an explicit counter-electrode and the solid lines are JDFT calculations (b) close-up view of $\langle\Phi(z)\rangle$ for JDFT calculations with $c=1.0\ \mathrm{M}$ (red) and $c=0.1\ \mathrm{M}$ (blue) and applied voltage ${\cal E}=-1.09\ \mathrm{V}$ vs. PZC (almost indistinguishable in the previous plot) (c) Variation of $\langle\Phi(z)\rangle$ in JDFT monovalent electrolyte of $c=1.0\ \mathrm{M}$ with ${\cal E}=\{-1.09,-0.55,0.0,0.55,1.09\}\ \mathrm{V}$ vs. PZC.}
\label{Figure3}
\end{figure}

Figure~\ref{Figure3}(a) contrasts the potential profiles resulting from the aforementioned approaches in actual calculations of a Pt(111) electrode surface.  Figure~\ref{Figure3}(a) displays the microscopic local electron potential energy function $\langle \phi(z)\rangle$ for a surface at applied voltage ${\cal E}=-1.09 V$ vs PZC which corresponds to a charge of $\sigma=-18\ \mathrm{\mu C/cm^2}$. The screened electron potentials generated by solution of Eq. \ref{mPB} at two different ionic strengths ($c=1.0$~M and $c=0.1$~M) are compared to potential profiles for a similarly charged surface in vacuum, with the net charge in the system neutralized either by imposing a uniform background charge or by placing an oppositely charged counter electrode at one Debye screening length from the metal surface. The two charge-compensated vacuum calculations clearly do not correspond to the electrochemical behavior, with far wider potential variations than expected.  Figure~\ref{Figure3}(b) shows a detailed view of the macroscopic {\em electrostatic} potential $\langle \Phi(z)\rangle$ for the same charged surface (obtained by subtracting the microscopic {\em electron} potential of the neutral surface and switching the sign to reflect electrochemical convention) for the JDFT calculated charged surface at two different ionic strengths.  The charge-compensated vacuum calculations would be off the scale of this figure, while the macroscopic electrostatic potential for the JDFT calculations obtains the value of the applied potential within the electrode and then approaches a well-established reference value of zero with the correct asymptotic behavior in the fluid region.

\subsection{Electrochemical double layer structure}

The Gouy-Chapman-Stern model, described in Section III(C), offers a well-known prediction of the structure of the electrochemical double layer, to which the potentials from our model correspond precisely.  The electrostatic potential profiles in the standard electrochemical picture include an initial, capacitor-like linear drop in $\langle\Phi(z)\rangle$ due to the outer Helmholtz layer (the Stern region), followed by a characteristic exponential decay to zero deep in the fluid (the diffuse Gouy-Chapman region). Our model naturally captures this behavior as a result of (a) the localization of the dielectric response and screening to the liquid region as described by $N_{lq}(r)$ through Eq. (\ref{eps-kapp}) and (b) the separation between the fluid and regions of high explicit electron density $n(r)$ through the definition of $N_{lq}(r)\equiv N_{lq}\left(n(r)\right)$ via Eq. (\ref{Nl}).  Both the Stern and Gouy-Chapman regions are clearly evident in Figures~\ref{Figure3}(b,c). We find the dielectric constant transition region appearing in Figure~\ref{Figure2}(b), approximately the width of a water molecule, to be essential to the accurate reproduction of the double layer structure.  The potentials for charged surfaces in Figures~\ref{Figure3}(b,c) first show a linear decay in the region $0~$\AA$<z-z_{Pt}<\Delta$, corresponding to the ``gap'' between the end of the surface electron distribution ($z_{Pt}$) and the beginning of the fluid region, precisely the behavior we should expect in the Stern region. For a Pt(111) surface at applied voltage -1.09~V vs. PZC, $\Delta=0.6~$\AA,
but the width of this gap is voltage-dependent (as shown in Figure~\ref{Figure4}(b)) and also varies with metal and crystal face. After the gap region, for $\Delta<z-z_{Pt}<\Delta+\gamma$ (where $\gamma=0.6$ as in \ref{Nl}), the dielectric constant in Figure~\ref{Figure2}(b) changes rapidly from about 10 to the bulk value $\epsilon_b\sim 80$, defining a transition region which ensures that no significant diffuse decay in the potential occurs until beyond the outer Helmholtz layer, thereby allowing proper formation of the diffuse Gouy-Chapman region for $z-z_{Pt}>\Delta+\gamma$.  We emphasize that we have not added these phenomena into our calculations {\em a posteriori}, but that they occur naturally as a consequence of our microscopic, albeit approximate, {\em ab initio} approach.

\subsection{Charging of surfaces with electrode potential}

\begin{figure}

\centering
\subfloat[]{\label{fig:4a}\includegraphics[width=8.6cm]{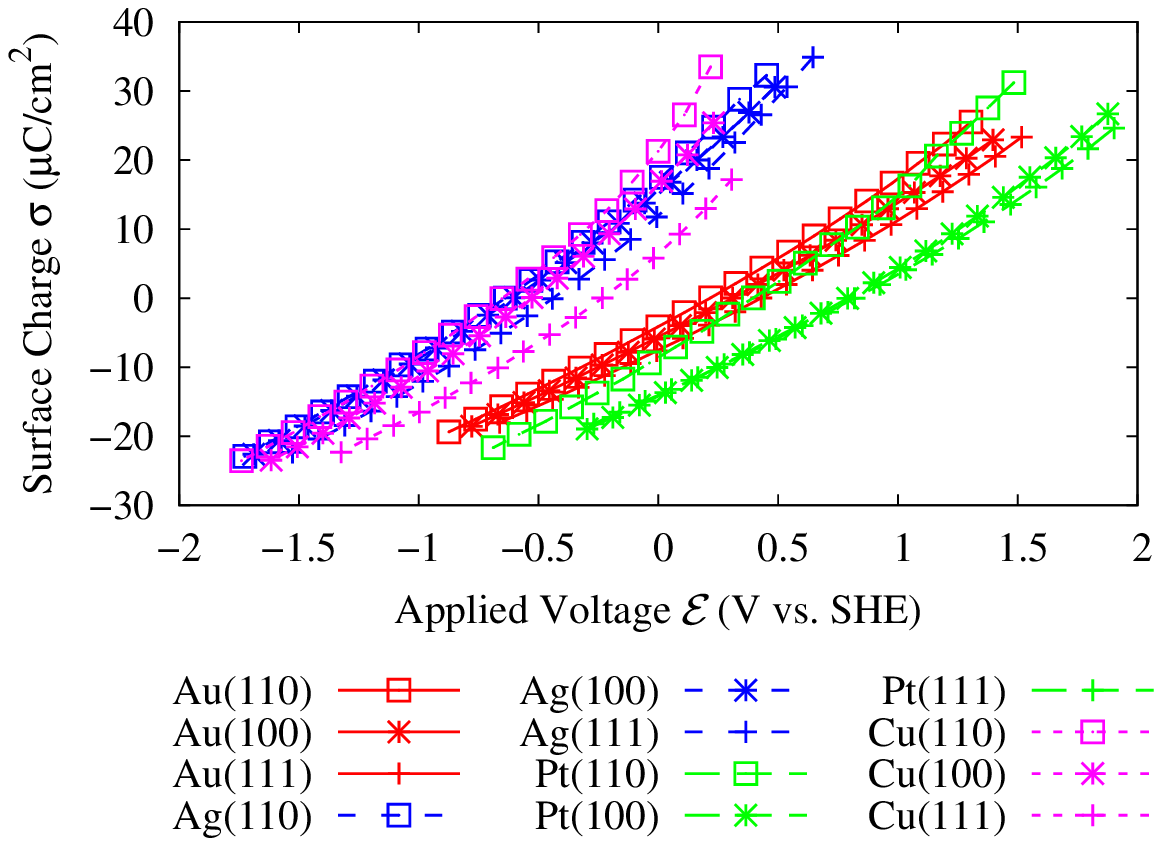}}

\subfloat[]{\label{fig:4b}\includegraphics[width=8.6cm]{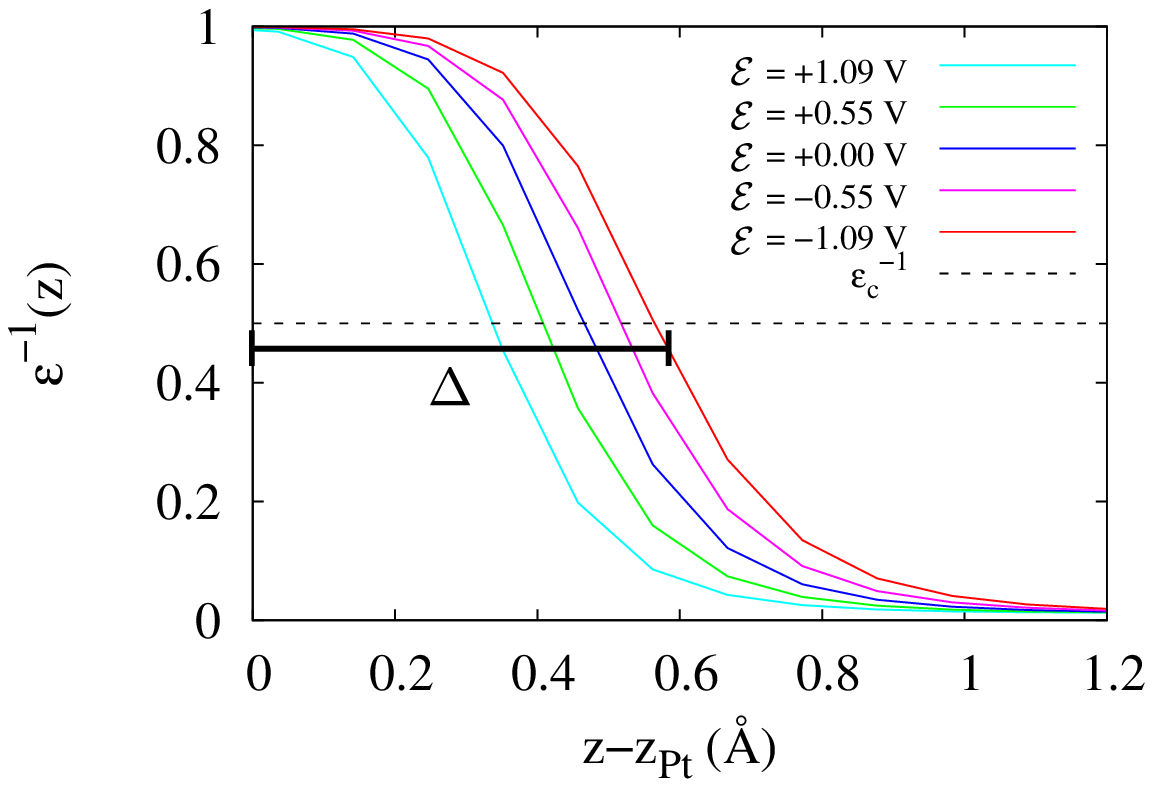}}

\subfloat[]{\label{fig:4c}\includegraphics[width=8.6cm]{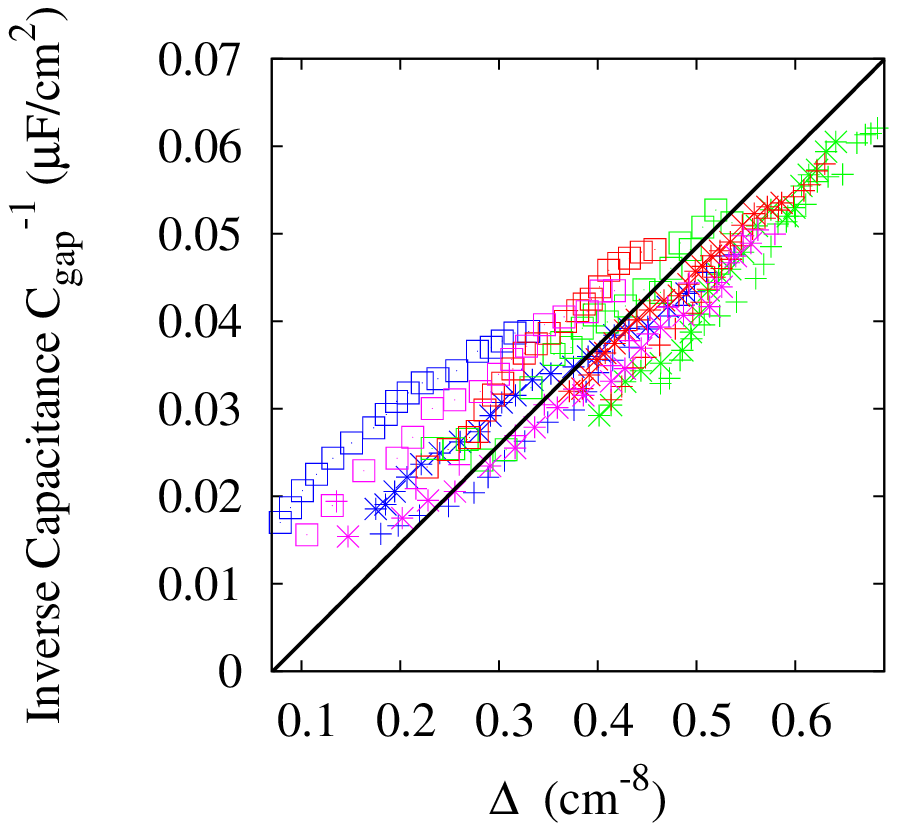}}

\caption{(a) Surface charge  $\sigma$ as a function of applied voltage ${\cal E}$ for a series of transition metal surfaces in an electrolyte of monovalent ionic strength $c=1.0\ \mathrm{M}$ (b) Inverse dielectric constant $\epsilon^{-1}$ as a function of distance from a Pt(111) surface for multiple values of applied voltage (c) Inverse gap capacitance ${\cal C}_{\Delta}^{-1}$ as a function of the distance from the metal surface at which the fluid begins $\Delta$. The solid line indicates the best fit to the data with slope constrained to $\epsilon_0^{-1}$}

\label{Figure4}
\end{figure}

To explore the effects of electrode potential on the surface charge and electronic structure, Figure~\ref{Figure4}(a) shows the surface charge $\sigma$ as a function of potential ${\cal E}$ for a series of transition metal surfaces for an electrolyte of monovalent ionic strength $ c=1.0\ \mathrm{M}$, without adsorption of ions to the surface.  We find the average double layer capacitance of the Pt(111) surface -- the slope of the corresponding $\sigma-{\cal E}$ curve in Figure~\ref{Figure4}(a) -- to be ${\cal C}$=19 $\mathrm{\mu F/cm^2}$, in excellent agreement with the experimental value of 20 $\mathrm{\mu F/cm^2}$.\cite{expCap}  Indeed, we find that a significant fraction of our total capacitance is due to dielectric and screening effects in the fluid; this agreement again supports our model for the electrolyte. The remainder is associated with the ``quantum capacitance'' or density of states ${\cal C}_{\mbox{DOS}}$ of the surface slab in our supercell calculations.

Closer inspection of the charge versus potential data reveals that the slope is not quite constant as a function of voltage. Indeed, taking the numerical derivatives of the curves in Figure~\ref{Figure4}(a) yields values for the differential capacitance that exhibit an approximately linear dependence on voltage. This voltage-dependence constrasts with studies performed using a different technique to produce voltage-independent constant values for the capacitance\cite{Norskov08}, which not only was limited to producing a constant value for the capacitance, but also required computationally demanding thermodynamic sampling to model the fluid.

To understand the origin of the above voltage-dependence of the capacitance, we employ the series model for differential capacitance in (\ref{Capseries}) and (\ref{CapFl}), in which the total capacitance per unit area ${\cal C}$ is modeled as a series combination of the capacitance associated with the density of states of the metal, a Stern capacitance (${\cal C}_{\Delta}$) across a gap of width $\Delta$, and the (constant) Gouy-Chapman capacitance associated with the inverse screening length $\kappa$.  We can then extract the ``gap'' capacitance as
\begin{flalign}
\frac{\Delta}{\epsilon_0}\sim{\cal C}_{\Delta}^{-1} \equiv {\cal C}^{-1}-{\cal C}_{\mbox{DOS}}^{-1}-\frac{\kappa^{-1}}{\epsilon_b\epsilon_0}.
\label{Cdelta}
\end{flalign}
To verify that the voltage-dependence of this contribution indeed correlates to changes in the gap associated with the Stern layer, we make an independent definition of the width of the gap as $\Delta\equiv z_c-z_{Metal}$, where $z_c$ represents the location where the presence of our model fluid becomes significant and $z_{Metal}$ represents the location of the surface of the metal. Specifically, we define $z_c$ as the point where the planar average of the inverse dielectric constant has fallen by half from its value in the electrode (as in Figure~\ref{Figure4}(b)) since the polarization of the fluid becomes significant when $\langle\epsilon^{-1}(z_c)\rangle<0.5$.  We determine $z_{Metal}$ from the covalent radii of the metal surface atoms, but note that the specific choice of $z_{Metal}$ is unimportant in the analysis to follow.

Figure~\ref{Figure4}(c) correlates the inverse gap capacitance ${\cal
  C}_{\Delta}^{-1}$ from the right hand side of Eq.~\ref{Cdelta} with
the values of $\Delta$ defined as in the previous paragraph.  There is
a striking linear trend with a slope within about ten percent of
$\epsilon_0^{-1}$, validating that the primary contribution to the
voltage-dependence of the differential capacitance within this model
is from changes in the gap between the fluid surface and where the
dielectric screening begins.  The ultimate origin of this effect
within the present approximation (in which the dielectric constant is
determined by the electron density through Eq.~\ref{eps-kapp}) can be
traced to the increase in surface electrons with decreasing applied
potential, which moves the location of the fluid transition further
away from the metal surface. In fact, the experimentally determined
capacitance of Pt(111) due to {\em only} the double layer\cite{expCap}
(after subtracting the effects of counter-ion adsorption) has a
voltage-dependence quite similar to our prediction. Since the distance
of closest approach of the fluid to the metal surface is determined by
van der Waals interactions and the addition of more electrons could
indeed strengthen the repulsion, the qualitative voltage-dependence of
the ``double-layer'' capacitance even at this simple level of
approximation may indeed be capturing some aspects of the underlying physics.

The double-layer capacitance notwithstanding, in physical systems the
total capacitance is dominated by the effects of adsorption of
counter-ions, and so the qualitative voltage-dependence of the
capacitance at this simple level of approximation has limited
practical relevance. Nonetheless, it is an important feature of the
electrochemical interface for those modified Poisson-Boltzmann approaches in which the
cavities are determined by contours of the electron density.  Future
work in this area could capture the ``ion-adsorption'' portion of the
capacitance either by including explicit counter-ions within the electronic structure portion of the calculation or by
choosing a classical fluid functional that includes a microscopic description
of the counter-ions.

\subsection{Potentials of Zero Charge and reference to Standard Hydrogen Electrode}
 
\begin{figure}
\begin{center}
\subfloat[LDA]{\label{fig:5a}\includegraphics[width=6.3cm]{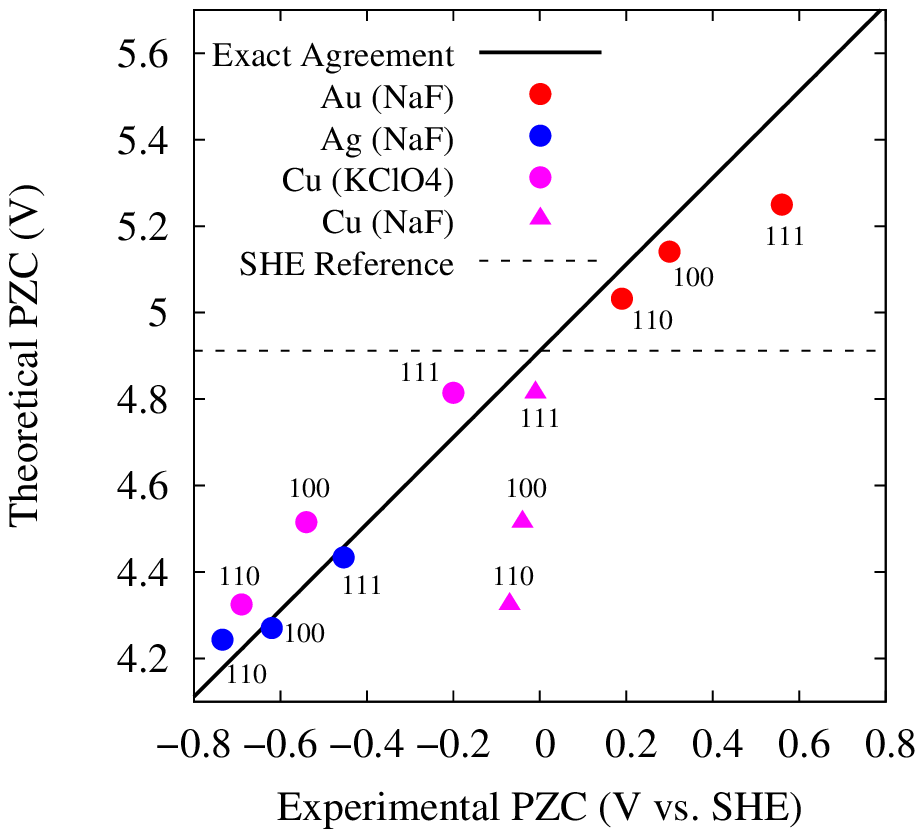}}

\subfloat[GGA]{\label{fig:5b}\includegraphics[width=6.5cm]{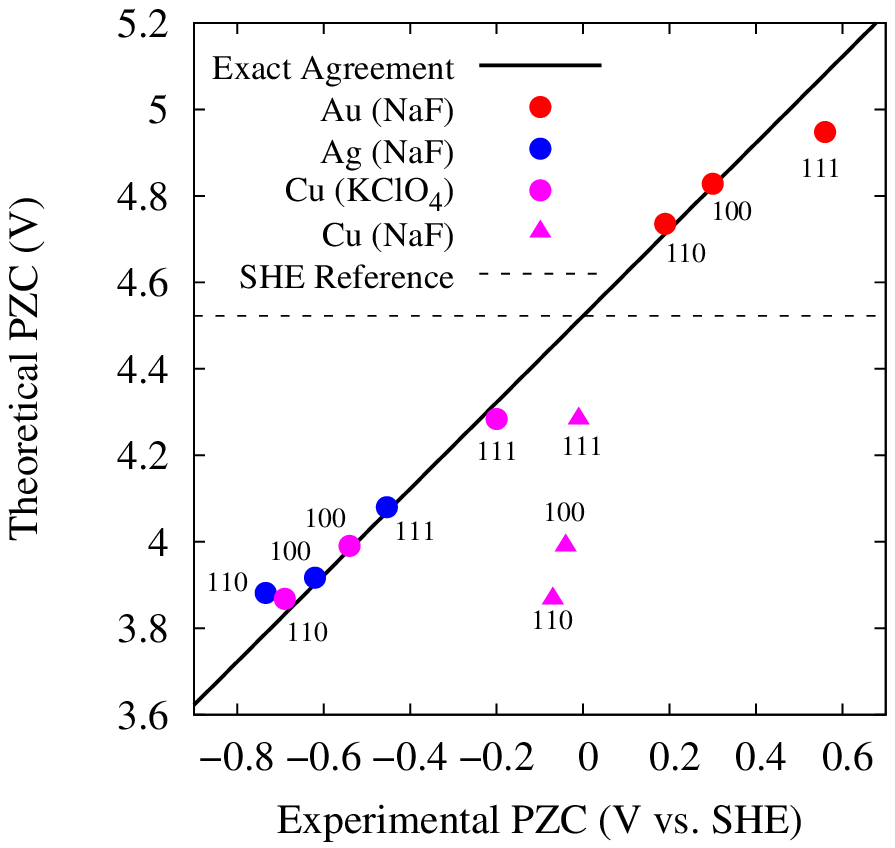}}

\subfloat[Comparison to Work Functions]{\label{fig:5c}\includegraphics[width=8.6cm]{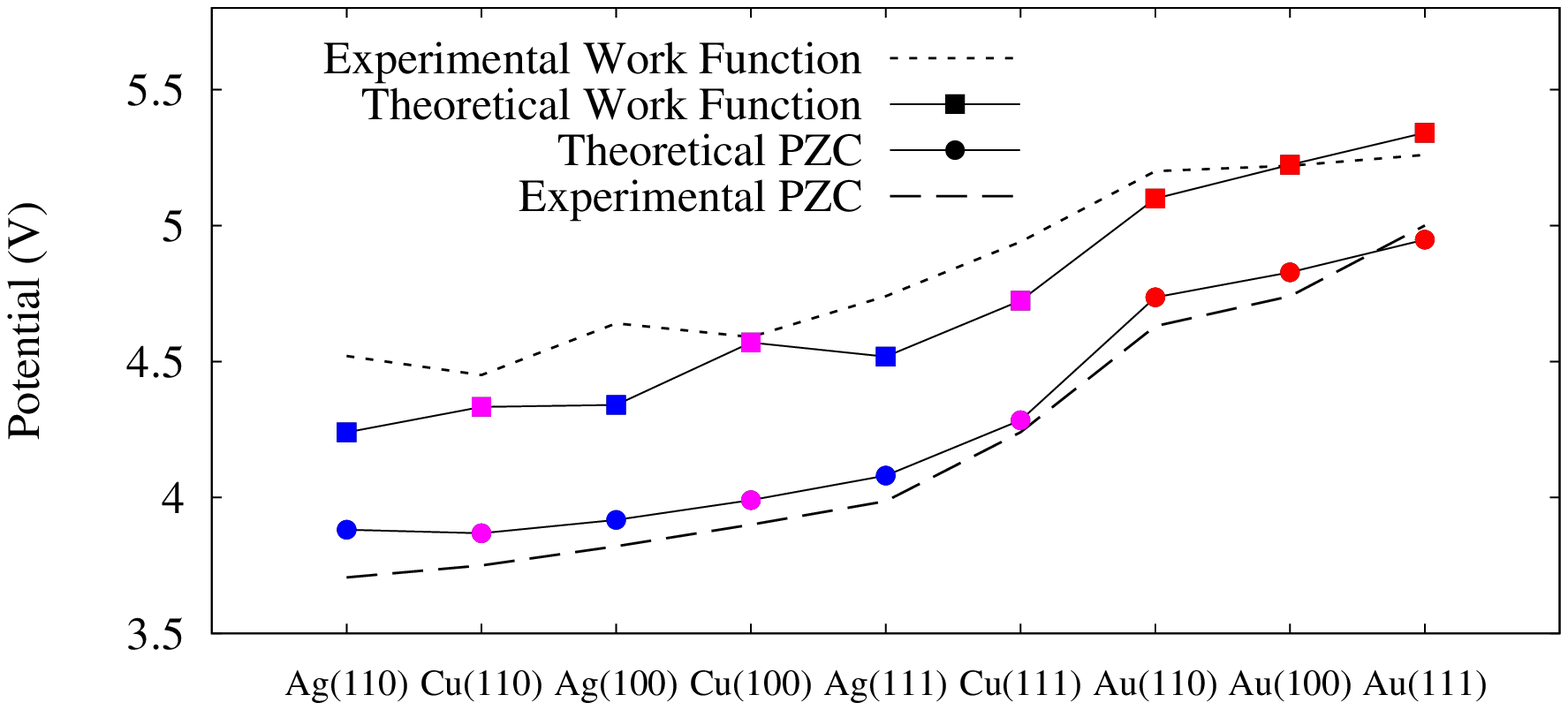}}
\end{center}
\caption{Comparisons of {\em ab initio} predictions and experimental data\cite{Trasatti} for potentials of zero free charge (PZCs) and vacuum work functions: (a) {\em ab initio} LDA predictions versus experimental PZC relative to SHE, (b) {\em ab initio} GGA predictions versus experimental PZC relative to SHE, (c) {\em ab initio} GGA vacuum work functions (solid line with squares) and PZC's (solid line with circles), experimental work functions (dotted line) and PZC's (dashed line) versus vacuum for the same series of surfaces.  Best linear fits with unit slope (dark diagonal solid lines in (a) and (b)).}
\label{Figure5}
\end{figure}

To connect our potential scale (relative to an electron solvated in our model fluid) to a standard potential scale employed in the literature and to confirm the reliability of our model, Figures~\ref{Figure5}(a,b) show our {\em ab initio} predictions for potentials of zero charge versus experimental values relative to the standard hydrogen electrode (SHE).\cite{Trasatti} Within both the local density (LDA)\cite{Kohn-Sham} and generalized gradient (GGA)\cite{PW91} approximations to the electronic exchange-correlation energy, we have calculated the potentials of zero charge for various crystalline surfaces of Ag, Au, and Cu, three commonly studied metals. We performed a least-squares linear fit to the intercept of our data, leaving the slope fixed at unity.  (Note that the experimental data for Cu in $\mathrm{NaF}$ electrolyte were not included in the fit, due to concerns discussed below.) The excellent agreement between our results (with a constant offset) and the experimental data indicates that joint density-functional theory accurately predicts trends in potentials of zero charge, and encourages us that it can establish oxidation and reduction potentials in the future.  The improved agreement of GGA (rms error: 0.058 V) over LDA (rms error: 0.108 V) underscores the importance of gradient-corrections to this type of surface calculation.  

The strong linear correlation with unit slope between the theoretical
and experimental data in Figures~\ref{Figure5}(a,b) indicates that the
simplified Poisson-Boltzmann approach reproduces potentials of zero
charge well relative to some absolute reference. The single parameter
in the fit for each of the two panels (namely, the vertical intercepts of
each fit line) establishes the absolute relationship between
our zero of potential (implicit in each set of theoretical results) and the
zero of potential on the standard hydrogen-electrode scale (implicit
in the experimental data).  Specifically, we find that our zero of
potential sits at -4.91~V relative to the SHE for LDA and -4.52~V
relative to the SHE for GGA. Intriguingly, these values are close to the experimentally determined
location of {\em vacuum} relative to the the standard hydrogen
electrode reference (-4.44~V)\cite{Trasatti}; in fact, the GGA reference value is within a tenth of a volt. This apparent alignment is not altogether surprising due to the following argument: (1) our method measures the
difference in energy between an electron in the electrode and an
electron solvated deep in our model electrolyte, so that our
potentials of zero charge are measured relative to a “solvated”
electron reference; (2) the energy of a solvated electron relative to
vacuum {\em within the presentlt considered linearized Poisson-Boltzmann model} is
zero because this approximation includes only electrostatic effects;
and (3) because the calculated potentials of zero charge in the
figures are thus {\em relative to vacuum}, the difference between our
calculated results and the experimental results should represent the
constant difference between the vacuum and SHE references.

Consideration of the breakdown of the potential of zero charge into physically meaningful quantities explains the difference between the LDA and GGA results and elucidates the apparent success of the rather simple modified Poisson-Boltzmann approach in predicting PZC's.  Transferring an electron from a metal
surface to a reference electrode requires, first, removal of the
electron from the surface and, then, transport of the electron through the relevant interfacial layers of the liquid. The energy associated with the fomer process is the work function, and the energy associated with the latter relates to the intrinsic dipole of the liquid-metal interface. As is well-known, there is an approximately constant
shift between the predictions of the LDA and GGA exchange and
correlation functionals for work functions of metals.  In fact, Fall
{\em et al.} report that GGA metal work functions are approximately
0.4 V lower than the LDA work functions,\cite{Fall2000} corresponding
well to the differences we find between the vertical intercepts of
Figures~\ref{Figure5}(a,b). Next, to aid consideration of the
intrinsic dipole of the interface, Figure~\ref{Figure5}(c) explicitly
compares our predictions for work function with our predictions for
potential of zero charge, including also the corresponding
experimental data for both quantities. (To place all values on a
consistent scale of potential, which we choose to be vacuum, we have
added the experimentally determined 4.44~V difference between SHE and vacuum
to the experimental PZC's.) The data in Figure~\ref{Figure5}(c) suggest that the vacuum work
functions are harder to predict than potentials of zero charge,
possibly due to difficulty determining the value of the reference
potential in the vacuum region, an issue not present in our fluid
calculations due to the screening in Eq. \ref{mPB}. The figure also
indicates an approximately constant shift from vacuum work function to
potential of zero charge, suggestive of a roughly constant interfacial
dipole for each of the metal surfaces.  However, the shift is not exactly
constant: both the experimental and theoretical data exhibit
significant fluctuations (on the order of 0.1 V) in the shift between
work function and PZC from one metal
surface to another. Because the PZCs are determined to within a
significantly smaller level of fluctuation (0.06 V), these data
indicate that the Poisson-Boltzmann model captures not merely a
constant interfacial dipole, but also a significant fraction of the
fluctuation in this dipole from surface to surface.

We note that Tripkovic {\em et al.} have also calculated the
potentials of zero charge for transition metal
surfaces.\cite{Rossmeisl11} However, that approach requires
calculation of several layers of explicit water within the electronic
structure portion of the calculation, and those authors find
the resulting potentials of zero charge to be dependent on the exact
structure chosen for the water layers. However, while differing orientations of
water molecules at the interface may result in significant local fluctuations in the
instantaneous PZC, the experimentally measured potential of zero
charge is a temporal and spatial thermodynamic average over all liquid
electrolyte configurations rather than the value from any single
configuration. Direct comparison to experimental potentials of zero
charge therefore should involve calculation of a thermodynamic
average.

As a matter of principle, derivatives of the free energy (which the
JDFT framework provides directly) yield thermodynamic averages.
Therefore, an exact free-energy functional would predict the exact,
thermodynamically averaged
potential of zero charge, and classical liquid functionals, which capture more
microscopic details of the equilibrium liquid configuration\cite{Lischner10} than the present model, would be an ideal choice for future in-depth studies.  Indeed, such functionals are capable of capturing the relevant electrostatic
effects even when a single configuration of water molecules dominates the thermodynamic
average.  (In such cases, minimization of the free-energy functional
results in localized site densities $N_\alpha(r)$ representing the
dominant liquid configuration.)
Of course, in cases of actual charge-transfer reactions between the
surface and the liquid, the (relatively few molecules) involved in the
actual transfer must be included within the explicit electronic density-functional theory, whereas the other electrolyte molecules may still be handled accurately within the more computationally efficient liquid density-functional theory.

There is also reason to be sanguine regarding the ability of the modified Poisson-Boltzmann approximation pursued in this
work to capture interfacial dipole effects.  The macroscopic dielectric constant contained within the present model describes primarily the orientational polarizability of water, so that the liquid bound charges resulting from the minimization of the free energy should reflect the most dominant configurations of water molecules in the thermodynamic average, even if only a single
configuration dominates.  On the electrode side, the image charges corresponding to the bound charge also naturally appear, as a consequence of both the electrostatic coupling in our model and the metallic nature of the surface described within electronic density-functional theory.  From an optimistic perspective, it is
quite possible that a significant portion of the electrostatics of the
surface dipole would be captured even at the simplified level of a
Poisson-Boltzmann description.  Ultimately, quantification of how much of the effect is captured may only be verified by comparison to experiment.  For the systems so far considered, the excellent {\em a priori} agreement between experimental measurements and our theoretical predictions indicates that the relevant effects are indeed
captured quite well.  It appears that even a simple continuum model (which only accounts for the effects of bound charge at the interface and the corresponding image charges within the metal) can predict accurately key electrochemical observables such as potential of zero charge.  Certainly, for more detailed future studies, we would recommend exploring the performance of more explicit functionals.  However, the apparent accuracy and computational simplicity of the current Poisson-Boltzmann approach render it well-suited for high-throughput studies of electrochemical behavior as a function of electrode potential.

As a further example of the utility of the Poisson-Boltzmann approach, the potential of zero charge calculation for copper illustrates how this theory can be used as a highly controlled {\em in-situ} probe of electrochemical systems, with the ability to isolate physical effects which are not possible to separate in the experiment.  Specifically, in Figures~\ref{Figure5} (a) and (b), for copper there are experimental values for two different electrolytes, NaF and $\mathrm{KClO_{4}}$, which are both claimed to be noninteracting with the metal surface.\cite{Trasatti} Clearly, our theoretical values, which correspond to potentials of zero {\em free} charge without adsorption of or chemical reaction with ions from the electrolyte, agree more favorably with experimental data for the Cu surface in $\mathrm{KClO_{4}}$ than for the NaF electrolyte. Our results suggest that future experimental exploration is warranted to investigate potential interactions between the NaF electrolyte and the copper surfaces, or into other possible causes of the discrepancy in potentials of zero charge. Perhaps some polycrystalline impurities caused the experimental potentials of zero charge of the supposed single-crystalline faces to become much more similar than our calculations and the $\mathrm{KClO_{4}}$  data indicate they should be. {\em Ab initio} calculations offer an avenue to study each of these potential causes independently and to elucidate the mechanisms underlying the apparent experimental disagreement.

Finally, although potentials of zero charge are quite readily observed in experiments for less reactive metals such as silver and gold, measurement of the potential of zero charge for platinum can be difficult because platinum is easily contaminated by adsorbates.  For this reason, more convoluted methods are employed to determine an experimental value for the potential of zero charge for platinum.  For instance, one may turn to ultra-high vacuum methods, where, by definition, no molecules are adsorbed on the surface, and one may then attempt to estimate the effect of the solution on the potential of zero free charge. \cite{MichaelWeaver}  Alternately, one may employ cyclic voltammograms to estimate the charge due to adsorbates and then extrapolate the potential of zero free charge. \cite{Cuesta}  Our {\em ab initio} method, however, gives the values for non-contaminated potentials of zero free charge directly, provided we establish the relation of our zero reference potential relative to that of the standard hydrogen electrode, which we have done above.  

For uncontaminated platinum, our method yields the potentials of zero free charge shown in Table \ref{PZCPt}. Compared to other references in the literature, the best agreement with our results is from an experiment which extrapolates the potential of zero charge from ultra-high vacuum, eliminating the effects of unknown adsorbates on the clean surface.\cite{MichaelWeaver} As with the results for Cu, the significantly better agreement of our calculations with this latter experimental approach suggests that perhaps future experiments which measure potential of zero charge should reconsider the effect of possible contaminants when extrapolating values for the potential of zero free charge.

\begin{table}[ht]
\caption{Platinum Potentials of Zero Free Charge (V~vs~SHE)}
\centering
\begin{tabular}{c c c c}
\hline\hline
 & (110) & (100) & (111) \\ [0.5ex]
\hline
LDA & 0.31 & 0.70 & 0.71 \\
GGA & 0.40 & 0.79 & 0.82 \\
\hline
\end{tabular}
\label{PZCPt}
\end{table}

\section{CONCLUSION}

In this work, we extend joint density-functional theory (JDFT) -- which combines liquid and electronic free-energy functionals into a single variational principle for a solvated quantum system -- to include ionic liquids. We describe the theoretical innovations and technical details required to implement this framework for study of the voltage-dependence of surface systems within standard electronic structure software.  We establish a connection to the fundamental electrochemistry of metallic surfaces, accurately predicting not only potentials of zero charge for a number of crystalline surfaces for various metals but also an independent value for the standard hydrogen electrode relative to vacuum.  Furthermore, we show how future innovations in free energy functionals could lead to even more accurate predictions, demonstrating the promise of the joint density-functional approach to predict experimental observables and capture subtle electrochemical behavior without the computational complexity required by molecular dynamics simulations. These advantages render joint density-functional theory an ideal choice for high-throughput screening calculations and other applications in materials design.

We have built extensively upon the framework of joint density-functional theory in the implicit solvent approximation,\cite{Petrosyan05} extending it to include charged ions in a liquid electrolyte.  Beginning with an implicit model for the fluid density $N_{lq}(r)$ in terms of the electronic density of the surface, $N_{lq}(r)=N_{lq}(n(r))$, we include an ionic screening length tied to the fluid density $N_{lq}(r)$ in the same way as done in previously successful models for the dielectric constant. We also solve a previously unrecognized difficulty by including model core electron densities within the surface to prevent artificial penetration of liquid density into the ionic cores, which lack electrons in typical pseudopotential treatments of the solid. Inclusion of this ionic screening allows us to provide a consistent zero reference of potential and to resolve many difficulties associated with net charges in periodic supercell calculations, thereby enabling study of electrochemical behavior as a function of applied voltage.

With the framework to include electrode potential within joint density-functional theory calculations thus in place, we then establish clear connections between microscopic computables and experimental observables.  We identify the electronic chemical potential of density-functional theory calculations with the applied voltage in electrochemical cells, and thereby extract a numerical value of 4.52~V (within the GGA exchange-correlation functional) for the value of the standard hydrogen electrode relative to vacuum, which compares quite favorably to the best-accepted experimental value of 4.44~V.\cite{Trasatti}  We also show that joint density-functional theory reproduces, {\em a priori}, the subtle voltage-dependent behaviors expected for a microscopic electrostatic potential within the Gouy-Chapman-Stern model and we extract potentials of zero free charge for a series of metals commonly studied in electrochemical contexts, often finding agreement with experimental values to within hundredths of volts.

This qualitatively correct prediction of electrochemical behavior and
encouraging agreement with experiment demonstrate the capabilities of
even a simple approximation within the joint density-functional theory
framework, and we expect future improvements to the free-energy
functional to be able to describe more complex electrochemical
phenomena. Future work should also generalize the approximate
functional to include nonlinear saturation effects in ionic screening
within the current modified Poisson-Boltzmann approach, with an
approach along the lines of other works. \cite{MarzariDabo}  In
electrochemical experiments, the differential capacitance of charged
metal surfaces often exhibits a minimum at the potential of zero
charge\cite{BardFaulkner} (not seen in the linear continuum theory),
and more advanced theories including such nonlinear effects should be
able to capture this more subtle behavior.  Additionally, recent
developments in classical density functionals for liquid water
\cite{Lischner10} now can be implemented to study electrochemical
systems. Such classical density functionals can be extended to include
realistic descriptions of ions and are capable of capturing other
essential behaviors of electrolyte fluids, including features in the
ion-ion and ion-water correlation functions due to differences in the
structure of the anion and the cation. \cite{Bazant} Finally, in
systems where electrochemical charge-transfer reactions are important
or where chemical bonds of the fluid molecules are expected to break,
the relatively few reactant molecules should be treated within the explicit
electronic structure portion of the calculation, with the remaining
vast majority of non-reacting molecules handled within the more computationally efficient liquid
density-functional theory.

With advances such as those described above, joint density-functional theory holds promise to become a useful and versatile complement to the toolbox of currently available techniques for first principles study of electrochemistry. Unlike {\em ab initio} molecular dynamics (or any other theory involving explicit water molecules), this computationally efficient theory is not prohibitive for larger system sizes.  In fact, as the system size grows, the fraction of calculation time spent solving the modified Poisson-Boltzmann equation actually decreases, meaning that for larger systems, the calculation is only nominally more expensive than calculations of the corresponding systems carried out in a vacuum environment. Also, because thermodynamic integration is not required, the joint density-functional theory approach yields equilibrium properties directly and has a clear advantage over molecular dynamics simulations for calculation of free energies.  Immediate applications include the study of molecules on metallic electrode surfaces as a function of applied potential and prediction of the basic properties of novel catalyst and catalyst support materials. These calculations could inform future materials design by offering an opportunity to screen novel complex oxides and intermetallic materials in the presence of the true electrochemical environment, thereby elucidating the fundamental physical processes underlying fuel cells and liquid-phase Graetzel solar cells.
 
\begin{acknowledgments}

The authors would like to acknowledge Ravishankar Sundararaman for modifying the software to streamline calculations at fixed voltage and Juan Feliu for providing the most up-to-date information regarding electrochemistry of single-crystalline metallic surfaces.

\vspace{5mm}

This material is based on work supported by:

\vspace{5mm}

The Energy Materials Center at Cornell, an Energy Frontier Research Center funded by the U.S. Department of Energy, Office of Science, Office of Basic Energy Science under Award Number DE-SC0001086.

\vspace{5mm}

The Cornell Integrative Graduate Education and Research Traineeship (IGERT) Program in the Nanoscale Control of Surfaces and Interfaces, supported by the National Science Foundation under NSF Award DGE-0654193, the Cornell Center for Materials Research, and Cornell University.

\vspace{5mm}

K. Letchworth-Weaver also acknowledges support from an National Science Foundation Graduate Research Fellowship.

\end{acknowledgments}


%

\appendix

\section{Implementation within standard electronic structure software}

Here we consider the issues which arise when implementing the above framework within a pre-existing electronic-structure code.  We will focus on software operating within the pseudopotential framework as this technique is commonly used for surface calculations.

Such pseudopotential calculations, for computational efficiency, include the nuclei and core electrons together as a unit and describe their combined effects on the valence electron system through effective, "pseudo"-potentials.  Two subtleties now arise.  First, because the pseudopotentials describe the long-range electrostatic interaction between the ionic cores and the electrons, the screening
of the long-range Coulomb part of the pseudopotentials by the electrolyte environment through (\ref{mPB}) must be handled properly. Second, because the calculated (valence) electron density $n(r)$ in the atomic core regions tends to be relatively low in pseudopotential calculations, our definition of the liquid density $N_{lq}(n)$ as a local function of the local electron density $n(r)$ through (\ref{Nl}) can lead to the unphysical presence of liquid within the atomic cores if precautions are not taken.

As a matter of notation specific to this appendix, we separate conceptually the valence electron density $n_v(r)$, calculated directly with the Kohn-Sham orbitals, from the missing contribution $n_c(r,\{R_I\})$ due to the core electrons, which clearly varies explicitly with the locations of the centers of the ions. By
including both electron and ionic (now, actually, valence-electron and ionic core) source terms, the new energy functional (\ref{ApproxFunc}) naturally provides electrolyte screening of all of the relevant fields.  The functional (\ref{ApproxFunc}) then becomes
\begin{flalign}
A[n(r),\phi(r)]&= A_{TXC}[n_v(r)]\nonumber\\
& +E_{C}[n_v(r),n_c(r,\{R_I\}),\phi(r)]
\nonumber\\& +U_{\mbox{ps}}[n_v(r),\{Z_I,R_I\}],
\end{flalign}
where  $A_{TXC}[n_v(r)]$ is the Kohn-Sham single-particle kinetic plus exchange-correlation energy, and
\begin{flalign}
E_{C}[n_v(r),&n_c(r,\{R_I\}),\phi(r)]=\nonumber\\ &\int
d^3r\{\phi(r)\left(n_v(r)-N(r,\{Z_I,R_I\})\right)\nonumber\\ &-
\frac{\epsilon(n_v(r)+n_c(r,\{R_I\}))}{8\pi}|\nabla
\phi(r)|^2\nonumber\\ &-\frac{\epsilon_b\,
\kappa^2(n_v(r)+n_c(r,\{R_I\}))}{8\pi}(\phi(r))^2\}
\label{EC}
\end{flalign}
represents the previously described electrostatic contributions to the total energy functional (with $N(r,\{Z_I,R_I\}) \equiv \sum_I Z_I \delta^{(3)}(r-R_I)$ representing point charges with the {\em ionic valences} $Z_I$), and the term
\begin{flalign}
U_{ps}[n_v(r),&{Z_I,R_I}]=\nonumber\\&\int{\left(\sum_I \Delta V^{(I)}_{ps}(r-R_I)\right)\, n_v(r)\ d^3r},
\label{Ups}
\end{flalign}
with
\begin{equation}
\Delta V^{(I)}_{ps}(r-R_I) \equiv V^{(I)}_{ps}(r-R_I)+{Z_I}G(r-R_I),
\label{deltaVps}
\end{equation}
represents the non-Coulombic components of the pseudopotential $V^{(I)}_{ps}(r)$, with $G(r)\equiv 1/|r|$ being the Coulombic Green's function associated with a unit point charge in free space.  Note that here and henceforth in this work $Z_I$ refers to the charges of the ionic pseudopotential cores and not the nuclear atomic numbers. Finally, Appendix~\ref{app:Numer} details how, for practical numerical reasons, we work not with mathematical point charges but rather with narrow charge distributions which we can resolve numerically.

The derivative of the functional  $A[n(r)]$ with respect to $n_v(r)$ at a point $r$ (which gives the local effective Kohn-Sham potential used for the electronic wavefunction minimization) must also be adjusted to include the new dielectric response and ionic screening terms,
\begin{flalign}
\frac{\partial}{\partial n_v(r)} A[n(r)]&=\frac{\partial}{\partial
n_v(r)} A_{TXC}[n_v(r)]\nonumber\\& +\sum_I \Delta V^{(I)}_{ps}(r-R_I)\nonumber\\
& +\phi(r)-\frac{1}{8\pi}(\frac{\partial\epsilon}{\partial n}
|\nabla\phi(r)|^2\nonumber\\& +\epsilon_b\frac{\partial \kappa^2}{\partial n} \,
\phi^2(r)).
\label{funcderiv}
\end{flalign}

Finally, Hellman-Feynman calculation of the forces on the atoms requires care because of the dependence of both the ionic core density  $N(r,\{Z_I,R_I\})$ and
the model core electron density $n_c(r,\{R_I\})$ on the ionic positions $R_I$.  The final result is
\begin{equation}
\nabla_{R_I} A=\nabla_{R_I}\,E_C + \nabla_{R_I}\,U_{ps}
\label{HellFeyn},
\end{equation}
where
\begin{flalign}
\nabla_{R_I}\,U_{ps}\,&=\int\left(\nabla_{R_I}\,\Delta V^{(I)}_{ps}(r-R_I)\right)n_v(r)\,d^3r\nonumber\\
\nabla_{R_I} E_C&=\nonumber\\
& -\frac{1}{8\pi}\int d^3r\left(
\frac{\partial\epsilon}{\partial
n}|\nabla\phi|^2\,+\epsilon_b\frac{\partial \kappa^2}{\partial
n} (\phi(r))^2\right) \nonumber \\
& \times  \nabla_{R_I} n_c(r,\{R_I\}) \nonumber\\
& -\int  d^3r\phi(r)Z_I\nabla_{R_I}\delta^{(3)}(r-R_I)
\end{flalign}

While some of these derivatives have simpler analytical forms than those described above, these forms render much simpler the numerical representation of the relevant quantities, particularly the Dirac-delta functions and pseudopotentials, as described in Appendix~\ref{app:Numer}.

\section{Numerical Details} \label{app:Numer}

The system-dependent modified Poisson-Boltzmann equation which appears in our calculations does not have a direct analytic solution in either Fourier or real space and, thus, requires a numerical solution such that we cannot employ analytic Dirac $\delta$ functions to represent the ion-core charges.  Instead, in our numerical calculations, we employ "smoothed" ion-core charge densities,
\begin{equation}
N^{(\sigma)}(r,\ \{Z_I,R_I\})=\sum_I Z_I \delta^{(\sigma)}(r-R_I)
\end{equation}
where $\delta^{(\sigma)}$
is a normalized, isotropic three-dimensional Gaussian of width $\sigma$ containing a single unit change. To the extent that the smoothed distributions do not overlap with each other or the fluid regions, the replacement of the point charges with these distributions will not affect the "Ewald" energy among the charged atomic cores or the dielectric screening effects.  In practice, we find good numerical solutions by employing a relatively narrow width determined by the spatial resolution of the calculation, so that $\sigma$ corresponds to a distance of 1.40 points on the Fourier grid.  (Specifically, this work employs a plane-wave energy cutoff of 30~H, so that $\sigma=0.168~$\AA.)  This choice of parameters ensures that there is little overlap among the Gaussians and between the Gaussians and the fluid, so that that this replacement has negligible effect on the screened interaction among the atomic cores.

These smoothed distributions, however, do overlap with the valence electrons, an effect which we must compensate.  We compensate this local effect {\em exactly} by replacing the point-charge response $G(r)$ in the modification of the pseudopotential (\ref{deltaVps}) with the response corresponding to the smoothed densities,
\begin{equation}
G^{(\sigma)}(r)\equiv \frac{\mbox{erf\,}(|r|/\sqrt{2\sigma^2})}{|r|}.
\end{equation}
We also represent the core-electron densities in Eq. (\ref{eps-kapp}) that prevent fluid penetration into the atomic cores, and whose form is thus not critical, with Gaussian distributions
\begin{equation} \label{eq:flcore}
n_c(r,\{R_I\})=C \sum_I \delta^{(r_{\mbox{c}})}(r-R_I).
\end{equation}
Alternately, for this density, one could use the core-electron density from the partial core correction from an appropriately designed pseudopotential. Because the role of $n_c$ in our framework is simply to prevent penetration of fluid into the ionic cores, the precise values of the norm $C$ and width  $r_{\mbox{c}}$ are not critical.  We find that the choice $C=0.3~$\AA$^{-3}$, $r_c=0.2~$\AA works well for this purpose for all the species in our calculations.

Finally, with the above definitions in place, we have taken care to make all replacements $\delta^3(r)\rightarrow \delta^{(\sigma)}(r)$, $G(r)\rightarrow G^{(\sigma)}(r)$ and $n_c(r,\{R_I\})=C\, \sum_I \delta^{(r_{\mbox{c}})}(r-R_I)$ in the appropriate places in the expressions for the total free energy, in the functional derivatives appearing in the effective Kohn-Sham potential (\ref{funcderiv}), and in the expressions for the Hellman-Feynman forces on the atoms (\ref{HellFeyn}).  These substitutions complete the numerical specification of the functionals employed in our calculations.

We find that standard electronic structure methods work well with our functionals.  The one equation whose solution requires new algorithms is the modified Poisson-Boltzmann equation, which, unlike the standard Poisson equation, does not have a direct analytic solution in Fourier space.  To solve this equation, we have, however, found a simple to implement, yet highly efficient preconditioned conjugate gradient algorithm.

The portions of the functional which depend on the potential field $\phi(r)$ appear in $E_C$ in Eq. (\ref{EC}), which is a quadratic functional whose maximum corresponds to the solution of the modified Poisson-Boltzmann equation, Eq. (\ref{mPB}), and whose quadratic kernel is
\begin{equation}
Q={(\nabla\cdot\epsilon\nabla-\epsilon_b\kappa^2})/{(4\pi)}.
\end{equation}
We chose to solve this equation in Fourier space, where the diagonal elements of this kernel have a very simple approximate form, which leads to the diagonal preconditioner,
\begin{equation}
K(G)=(\bar{\epsilon} G^2+\epsilon_b\bar{\kappa^2})^{-1},
\end{equation}
where  $\bar{\epsilon}=\frac{1}{\Omega}\int\epsilon(r)d^3r$ and $\bar{\kappa^2}=\frac{1}{\Omega}\int\kappa^2(r)d^3r$  are the average values of these parameters over the unit cell. This diagonal preconditioner completely ignores the spatial variation of the dielectric constant.  A more effective preconditioner   which may be obtained by building in this inhomogeneity -- is calculated by first multiplying by $\sqrt(K(G))$ (the square root of the diagonal preconditioner) in Fourier space, transforming to real space and dividing by $\epsilon(r)$, then returning to Fourier space and again multiplying by $\sqrt(K(G))$. This inhomogeneous preconditioner requires more time to evaluate for a single iteration than the diagonal preconditioner, but reduces the total number of iterations required significantly.

\end{document}